\documentclass[]{emulateapj}
\usepackage{apjfonts}
\newcommand{\be}{\begin{equation}}
\newcommand{\ee}{\end{equation}}
\newcommand{\Msun}{M_{\odot}}
\def\kms{\ {\rm km\, s}^{-1}}
\def\teff{\ T_{\rm eff}}
\def\lbol{\ L_{\rm bol}}
\def\lsun{\ L_\Sol}

\shortauthors{CONROY \& VAN DOKKUM}
\shorttitle{COUNTING LOW-MASS STARS IN INTEGRATED LIGHT}

\begin{document}

\title{Counting Low-Mass Stars in Integrated Light}

\author{Charlie Conroy\altaffilmark{1}
  \& Pieter van Dokkum \altaffilmark{2}}

\altaffiltext{1}{Harvard-Smithsonian Center for Astrophysics,
  Cambridge, MA, USA} 
\altaffiltext{2}{Department of Astrophysical
  Sciences, Yale University, New Haven, CT, USA}

\begin{abstract}

  Low-mass stars ($M\lesssim0.4\Msun$) are thought to comprise the
  bulk of the stellar mass of galaxies but they constitute only of
  order a percent of the bolometric luminosity of an old stellar
  population.  Directly estimating the number of low-mass stars from
  integrated flux measurements of old stellar systems is therefore
  possible but very challenging given the numerous variables that can
  affect the light at the percent level.  Here we present a new
  population synthesis model created specifically for the purpose of
  measuring the low-mass initial mass function (IMF) down to
  $\sim0.1\Msun$ for metal-rich stellar populations with ages in the
  range $3-13.5$ Gyr.  Our fiducial model is based on the synthesis of
  three separate isochrones and a combination of optical and near-IR
  empirical stellar libraries in order to produce integrated light
  spectra over the wavelength interval $0.35\mu m<\lambda<2.4\mu m$ at
  a resolving power of $R\approx2000$.  New synthetic stellar
  atmospheres and spectra have been computed in order to model the
  spectral variations due to changes in individual elemental
  abundances including C, N, Na, Mg, Si, Ca, Ti, Fe, and generic
  $\alpha$ elements. We demonstrate the power of combining blue
  spectral features with surface gravity-sensitive near-IR features in
  order to simultaneously constrain the low-mass IMF, stellar
  population age, metallicity, and abundance pattern from integrated
  light measurements.  Finally, we show that the {\it shape} of the
  low-mass IMF can also be directly constrained by employing a suite
  of surface gravity-sensitive spectral features, each of which is
  most sensitive to a particular mass interval.
 
\end{abstract}



\section{Introduction}
\label{s:intro}

In order to understand the formation and evolution of galaxies we must
understand their stellar content.  Attention is often directed at
constraining the history of star formation, metallicity, and dust
content of galaxies.  In such work a key assumption is made: that the
stellar initial mass function (IMF) is universal and equal to the IMF
measured in the solar neighborhood.  This assumption, and its
influence on the derived physical properties of galaxies, is arguably
the largest source of uncertainty in nearly all work related to the
stellar populations of galaxies.

The IMF cannot be easily measured in extragalactic systems (or even in
our own Galaxy) because the stellar luminosity-mass relation is
steeper than the IMF --- the former scales as $M^{3-4}$ on the main
sequence (and is much steeper along the giant branches) while the
latter scales as $M^{-2.3}$ or shallower in the solar neighborhood.
Barring gross variation in these scalings in other galaxies, they
imply that the light from any population of stars is dominated by the
most massive stars that are still alive.  The faint low-mass stars
(i.e., $M<0.4\Msun$), which probably dominate the total stellar mass
budget in galaxies, contribute only of order $1$\% to the integrated
light.

In early type galaxies the problem is compounded by the fact that the
stars that dominate the light, K and M giants, have a similar spectral
type as the stars that dominate the mass budget, the K and M dwarfs.
This implies that, to first order, the light from low mass stars looks
similar to the light that dominates the flux.  This is why it has
historically been so difficult to measure the low-mass end of the IMF
in early type galaxies.

Progress can be made by realizing that the spectra of M dwarfs and M
giants at the same effective temperature display subtle but important
differences.  One of the first detections of such a difference was by
\citet{Wing69} who discovered a very strong absorption feature at
$0.99\mu m$ in the spectrum of an M dwarf that is not seen in M
giants.  This feature, now known as the Wing-Ford band, is attributed
to absorption by FeH.  It is very strong in late M dwarfs, but is
absent in M giants.  Among dwarfs, it is also very sensitive to
effective temperature.

There are in fact a number of strong absorption features that vary
with surface gravity at fixed effective temperature.  These features
include NaI at $0.82\mu m$, $1.14\mu m$ and $2.21 \mu m$, CaI at
$1.98\mu m$, CaII near $0.86\mu m$, CO near $2.30\mu m$, and H$_2$O
lines that are numerous in the near-IR.  The NaI, CaI and H$_2$O lines
are strong in dwarfs and weak or absent in giants, while CaII and CO
are strong in giants and weak in dwarfs \citep[see e.g.,][]{Frogel78,
  Kleinmann86, Diaz89, Ivanov04, Rayner09}.

It was recognized early on that dwarf-sensitive and, to a lesser
degree, giant-sensitive features could be used to count the number of
low-mass stars in integrated light\footnote{By ``integrated light'' we
  mean the combined light from the entire stellar population, as
  opposed to resolved photometry of individual stars.}
\citep[e.g.,][]{Spinrad62, Spinrad72, Cohen78, Frogel78, Frogel80,
  Faber80, Carter86, Hardy88, Couture93, Vazdekis96, Schiavon97a,
  Schiavon97b, Schiavon00, Cenarro03}.  The fundamental difficulty
then, as now, was the separation of abundance effects from
giant-to-dwarf ratio effects.  Both effects can change the strength of
the gravity-sensitive lines.  Since the average metallicity of massive
elliptical galaxies is believed to be greater than our own Galaxy
\citep[e.g.,][]{Worthey92, Trager00, Thomas05}, the use of stellar
spectra gathered in the solar neighborhood for the interpretation of
metal-rich ellipticals was viewed with skepticism.  The collection of
M giant spectra in the Galactic bulge, which is more metal-rich than
the disk, was recognized to be an important step in understanding the
spectra of massive ellipticals \citep{Carter86, Frogel87}, but this
approach has not yet been as fruitful as initially hoped.

The early work on this topic suffered from additional uncertainties.
The lack of accurate stellar evolution calculations across the main
sequence and through advanced evolutionary phases implied that model
construction suffered from major uncertainties that were difficult to
quantify.  Another major limitation was the poor near-IR detector
technology that existed in the 1980s and 1990s.  This made it very
difficult to measure spectra at the sub-percent level precision
necessary to measure the low-mass IMF in integrated light.  In the
intervening decades both of these uncertainties have been
significantly reduced (though not eliminated).  In addition,
substantial progress has been made in understanding the response of
stellar spectra to elemental abundance variations
\citep[e.g.,][]{Tripicco95, Korn05, Serven05}.  All of these factors
imply that the time is ripe to reconsider the possibility of measuring
the IMF from integrated light spectra.

Recently we obtained spectra for eight massive galaxies in the Coma
and Virgo clusters with $\sigma>250\kms$ \citep{vanDokkum10}.  Thanks
both to the new CCDs installed on the Low Resolution and Imaging
Spectrometer (LRIS) mounted on the Keck I telescope, and the stacking
of these spectra in the restframe (thereby averaging out any residual
sky emission and absorption problems), we were able to obtain accurate
spectra at the $<0.5$\% level.  We combined these data with a
preliminary version of the population synthesis model described herein
to conclude that in these massive ellipticals the ratio of low to
solar mass stars is much larger than in the Galaxy.  The IMF in these
massive ellipticals appeared to be much more `bottom-heavy' compared
to the IMF in the Galaxy.

The primary uncertainty in that work was the importance of abundance
effects; the empirical stars used in our model are of approximately
solar metallicity, while the ellipticals have metallicities in excess
of solar and are $\alpha$-enhanced \citep[e.g.,][]{Worthey92,
  Trager00, Thomas05}.  In a follow-up to our initial results, we
obtained spectra for four globular clusters (GCs) in M31 that have
iron abundances and abundance patterns (i.e., $\alpha-$enhancement;
[$\alpha$/Fe]) comparable to the massive ellipticals
\citep{vanDokkum11}.  A direct comparison of the spectra of these two
populations revealed significant differences {\it only in the surface
  gravity-sensitive lines}.  This constituted a critical test of our
interpretation of the elliptical data because the M31 GCs have low
mass-to-light ratios \citep{Strader11}, and so could not have
bottom-heavy IMFs.  We therefore concluded that the ellipticals
contained proportionally many more low mass stars than the GCs, again
consistent with a dwarf-rich IMF in the massive ellipticals.

In the present work we present detailed population synthesis models
with the goal of extracting the low-mass contribution to the
integrated flux.  The models span a range in age, metallicity,
abundance patterns, IMFs, and cover the wavelength range $0.35\mu m <
\lambda < 2.4\mu m$ at a resolving power of $R\approx2000$.  An
important feature of our approach is the use of synthetic spectra to
gauge the sensitivity of the spectrum to individual elemental
abundance variations.  Our principle goal is to show that it is
possible to separate the effects of IMF, age, metallicity, and
abundance pattern, through the consideration of a variety of spectral
features across the optical and near-IR wavelength range.  These
models can be contrasted with models such as \citet{Bruzual03}, which
model spectral energy distributions at solar abundance patterns, and
models that consider the effect of age, metallicity, and abundance
patterns on the Lick index system \citep[e.g.,][]{Trager00, Thomas03,
  Schiavon07}.  The model presented herein is, to our knowledge, the
first to consider the response of the entire optical and near-IR
spectrum to variations in individual elemental abundance patterns.

In the following sections we describe the details of the model
($\S$\ref{s:model}), provide a brief comparison to data
($\S$\ref{s:data}), present our results ($\S$\ref{s:res}), and discuss
several implications and set our findings in a broader context
($\S$\ref{s:disc}).  A summary of our main conclusions is also
provided ($\S$\ref{s:sum}).

\section{Model Construction}
\label{s:model}

\subsection{Overview}
\label{s:over}

Our goal is to construct models for the integrated light spectrum of a
population of stars as a function of population age, metallicity,
abundance pattern, and IMF.  Two ingredients are necessary to
construct such models: stellar evolution calculations (i.e.,
isochrones), and stellar spectral libraries.  The isochrones dictate
which stars are to be included in the synthesis, depending on the age
and metallicity of the stellar population.  The isochrones also
determine the stellar mass of each star, given the physical parameters
of the star (effective temperature, $\teff$, and bolometric
luminosity, $\lbol$).  With the appropriate list of stars for a given
age and metallicity, and the stellar mass of each star, the integrated
light single stellar population (SSP) spectrum can be simply
constructed from the integral:
\noindent
\be
f(\lambda) = \int_{m_l}^{m_u(t)} s(\lambda,m)\,\phi(m)\,{\rm d}m,
\ee
\noindent
where the integral over stellar masses, $m$, ranges from the hydrogen
burning limit, $m_l$, to the most massive star still alive at time
$t$, $m_u(t)$.  Throughout this paper we adopt $m_l=0.08\Msun$.  In
the above equation, $f$ is the integrated spectrum, $s$ is the
spectrum of an individual star, and $\phi$ is the IMF.  We will
consider four IMFs in the present work: a \citet{Chabrier03} IMF,
which is a fit to the IMF of stars in the disk of the Milky Way, a
Salpeter IMF, which has a logarithmic slope of $x=2.3$
($x\equiv-$dln$\phi$/dlnm), a `bottom-heavy' IMF that has a logarithmic
slope of $x=3$, and a `bottom-light' IMF advocated by
\citet{vanDokkum08}.  The bottom-light IMF has a similar shape to the
Chabrier IMF except that the former rolls over at $\approx2\Msun$
while the latter rolls over at $\approx0.1\Msun$.

In the following sections we describe the isochrones, stellar
libraries, and synthesis techniques used to create integrated spectra
as a function of population age, metallicity, and IMF.

\begin{figure}[!t]
\center
\resizebox{3.5in}{!}{\includegraphics{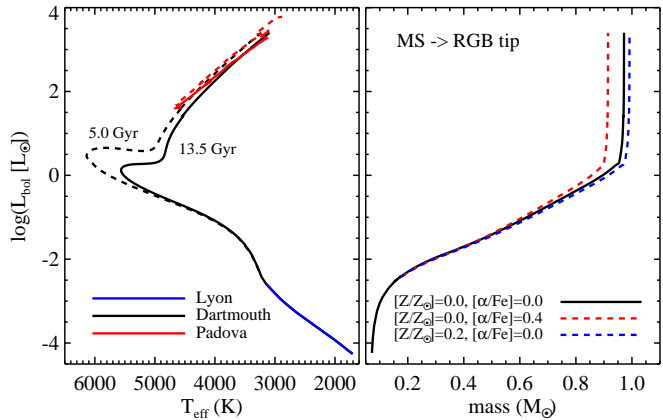}}
\caption{{\it Left panel:} HR diagram showing how stellar evolution
  calculations from three groups are stitched together to create an
  isochrone that spans all stages of stellar evolution (through the
  end of the AGB) for all masses.  Isochrones are shown for two ages
  as labeled in the panel. {\it Right panel:} Luminosity--mass
  relation as a function of metallicity.  Here only the
  Dartmouth-based isochrones are shown for simplicity.  Notice that
  the $L_{\rm bol}(M)$ relation is essentially insensitive to
  metallicity aside from the change in the turn-off mass.}
\vspace{0.1cm}
\label{fig:iso1}
\end{figure}

\subsection{Isochrones}
\label{s:iso}

\begin{figure}[!t]
\center
\resizebox{3.in}{!}{\includegraphics{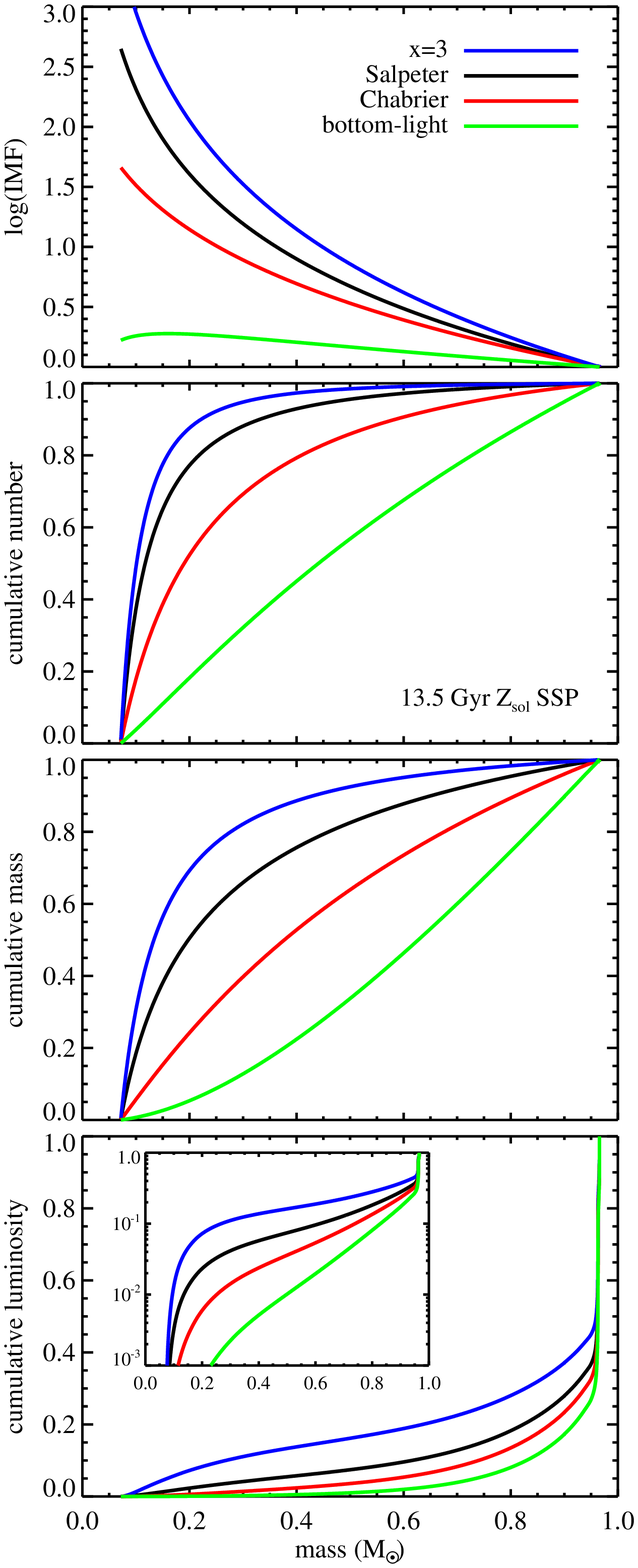}}
\caption{Cumulative contribution to the total number of stars, total
  stellar mass, and bolometric luminosity as a function of stellar
  mass.  Contributions are plotted as a fraction of the total.  The
  inset in the bottom panel shows the cumulative luminosity on a
  logarithmic scale.  Stellar remnants are not included in this
  figure.  Results are shown for three IMFs, for models with an age of
  13.5 Gyr and solar metallicity.  The small contribution from
  low-mass stars to the integrated luminosity, despite their large
  contribution to the total number of stars and the stellar mass, is a
  direct consequence of the $L-M$ relation shown in Figure
  \ref{fig:iso1}.}
\label{fig:fraciso}
\end{figure}

Accurate isochrones are required that extend from the hydrogen burning
limit to the end of the asymptotic giant branch (AGB) phase.
Unfortunately, no single set of publicly-available stellar evolution
calculations exists that meets these requirements.  We therefore need
to stitch together several separate evolutionary calculations.  For
the bulk of the main sequence and red giant branch (RGB) we use the
Dartmouth isochrones \citep{Dotter08b}, which have been shown to
reproduce very well the main sequence morphology of old star clusters
\citep{An09}.  These isochrones only extend to core helium ignition at
the tip of the RGB.  We therefore use the Padova isochrones
\citep{Marigo08} for horizontal branch (HB) and AGB
evolution\footnote{Due to different assumptions in the Dartmouth and
  Padova models, isochrones at a given age and metallicity do not join
  perfectly smoothly in mass between the tip of the RGB in the
  Dartmouth models and the HB in the Padova models.  In order to
  ensure a monotonic increase in mass along the combined isochrone, we
  have shifted the Padova masses so that they join smoothly with the
  end of the Dartmouth isochrones.  These shifts range from
  $10^{-3}\Msun$ at 13 Gyr to $0.05\Msun$ at 3 Gyr, and they have no
  effect on our results.}.  For $M<0.17\Msun$ we use the Lyon
isochrones \citep{Chabrier97, Baraffe98}.  The maximum age in the Lyon
models is 8 Gyr.  Since these are low mass stars that evolve little
over the range $3-14$ Gyr, we use their 8 Gyr models throughout.  The
precise value of 0.17$\Msun$ was chosen so that the Dartmouth
isochrones join smoothly with the Lyon isochrones in $\teff$ and
$\lbol$.

\begin{figure*}[!t]
\center
\resizebox{6.5in}{!}{\includegraphics{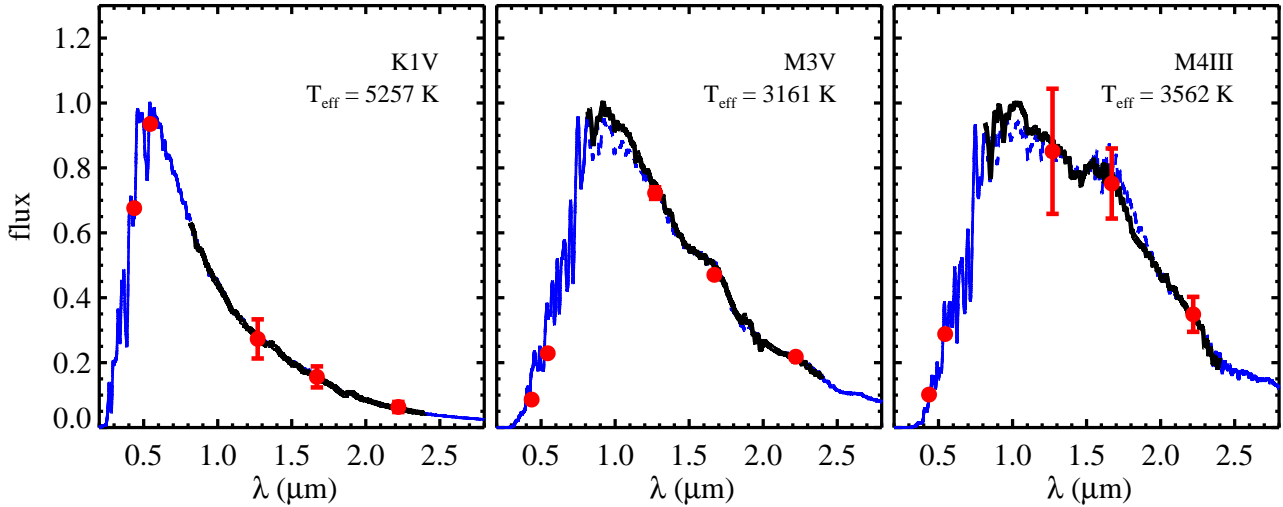}}
\vspace{0.4cm}
\caption{Spectral energy distributions for three stars from the IRTF
  library.  The IRTF spectra ({\it solid black lines}) are compared to
  optical and near-IR photometry ({\it symbols}) and our
  synthetic spectra ({\it blue lines}).  The synthetic spectra were
  selected to have the same $\teff$ and ${\rm log}(g)$ as the observed
  stars and were scaled to match the observed spectra at the blue and
  red ends of the data.  The theoretical spectra are dashed where they
  overlap in wavelength with the IRTF spectra, and solid where they
  are used to extrapolate the IRTF spectra.  The SEDs were normalized
  to unity and smoothed by 100\AA\, for display purposes.  Spectral
  types and estimated effective temperatures are shown in each panel.}
\label{fig:sed}
\end{figure*}

The switch to the Lyon models below $0.17\Msun$ is made because the
Dartmouth models apply the surface boundary condition at the
photosphere ($T=\teff$) rather than at the base of the atmosphere
($\tau=100$) as in the Lyon models.  The depth at which the surface
boundary condition is set is particularly important for the high
density, low temperature environment encountered in very low mass
stars.  Stellar interior codes are not ideally suited to compute the
physical conditions all the way to $T=\teff$.  We note in passing that
the Padova models are even less accurate at low masses than the
Dartmouth models, and they only extend to $0.15\Msun$, which is why we
use the Padova models only for advanced evolutionary phases.

In the left panel of Figure \ref{fig:iso1} we show the resulting
composite isochrone for a solar metallicity population with an age of
13.5 Gyr.  For comparison, we also show an isochrone for a 5 Gyr solar
metallicity population.  The right panel of the same figure shows the
$L-M$ relation for the main sequence and RGB.  In this panel we also
show the sensitivity of the $L-M$ relation to metallicity.  The
metallicity variations include a 0.2 dex increase in the total metal
abundance (Z), and a 0.4 dex increase in the $\alpha$-elements
([$\alpha$/Fe]).  Notice that while the main sequence turn-off mass
changes slightly, the main sequence $L-M$ relation is almost entirely
insensitive to these modest metallicity changes.

In the following we make use of the isochrones in two respects.
First, they are used to determine which stars to include in the
synthesis (e.g., stars that fall near the isochrone in the left panel
of Figure \ref{fig:iso1}).  Second, they are used to assign masses to
the stars based on $\lbol$.  Throughout this paper we will use the
same solar metallicity isochrones, even when synthesizing models with
different abundance patterns.  This simplification is motivated by two
facts: first, the $L-M$ relation is almost entirely insensitive to
metallicity over the range of metallicities we will consider (right
panel of Figure \ref{fig:iso1}).  This means that the masses assigned
to stars are robust against changes in metallicity.  Second, although
changes in metallicity induce changes in the temperatures of stars,
this effect is dwarfed by the much larger change in the spectra of
stars at fixed temperature when metallicity varies, especially for old
ages \citep[>5 Gyr; see][]{Coelho07, Schiavon07, LeeHC09}.

In Figure \ref{fig:fraciso} we show the IMFs considered herein and the
cumulative contribution to the total number of stars, total stellar
mass, and $\lbol$ as a function of stellar mass and IMF.  Stellar
remnants are not included in the figure (though they are included in
the total stellar mass budget in the rest of the paper).  This figure
summarizes the well-known fact that for standard IMFs low-mass stars
dominate the total stellar mass and number of stars, but contribute
only a small fraction of the total bolometric luminosity.

\subsection{Empirical spectral libraries}
\label{s:espec}

Empirical spectral libraries constitute the core feature of our model.
We make use of two separate libraries, the MILES library, which covers
the wavelength range $0.35\mu m<\lambda<0.74\mu m$
\citep{Sanchez-Blazquez06}, and the IRTF library of cool stars, which
covers the wavelength range $0.81\mu m<\lambda<2.4\mu m$
\citep{Cushing05, Rayner09}.

We select stars from the main IRTF library that are not classified as
chemically peculiar, have measured parallaxes (as obtained through
SIMBAD) and that belong to luminosity class III or V.  This leaves us
with 91 stars ranging in spectral type from F0 to M9.  The IRTF
library includes flux calibrated spectra and $BVJHK$ photometry for
each star.  Absolute flux calibration is provided by the $JHK$
photometry.  The spectral resolution is $R=2000$, and the
signal-to-noise ratio varies from $\sim100-1000$ with a median at
$1\mu m$ of 700.  We do not attempt to correct for dust extinction
because the $E(B-V)$ values provided in the IRTF library are
consistent with zero within the measurement uncertainties.  In
addition, the average distance to the stars in our sample is only 32
pc, suggesting that the spectra suffer from minimal extinction.

Effective temperatures are estimated for each IRTF star via the
standard technique of using calibrated effective temperature - color
relations.  The $\teff-(V-K)$ relation for giants was adopted from
\citet{Alonso99} for F0-K5 giants, \citet{Ridgway80} for K5-M6 giants,
and from \citet{Perrin98} for M6-M8 giants.  For dwarfs we adopt the
new $\teff$ scale from \citet{Casagrande08,Casagrande10}.  The $\teff$
scale of Casagrande et al. is consistent with earlier relations from
\citet{Alonso96} and \citet{Ramirez05}, modulo $\sim100$ K offsets
that can be attributed to differences in the photometric absolute
calibration adopted by different authors.

Bolometric luminosities were estimated for each IRTF star by
extrapolating the observed spectrum both blueward and redward with a
spectrum from our synthetic library (described in the next section) at
the same $\teff$.  Figure \ref{fig:sed} illustrates this procedure.
In each panel we show an IRTF spectrum, the accompanying $BVJHK$
photometry, and the synthetic spectrum used to extrapolate the IRTF
spectrum both blueward and redward.  Integrating over the full
wavelength range then yields an estimate of $\lbol$.  The wide
wavelength range of the IRTF spectra ensures that the bulk of the
luminosity is directly measured for late-type stars.  For example, the
IRTF spectra samples $70-85$\% of the bolometric luminosity for M-type
stars, $40-70$\% for G and K-type stars, and $20-40$\% for F-type
stars.  The $B$ and $V$ photometry provide an important consistency
check for our blueward extrapolation, as is evident from Figure
\ref{fig:sed}.

\begin{figure}[!t]
\center
\resizebox{3.5in}{!}{\includegraphics{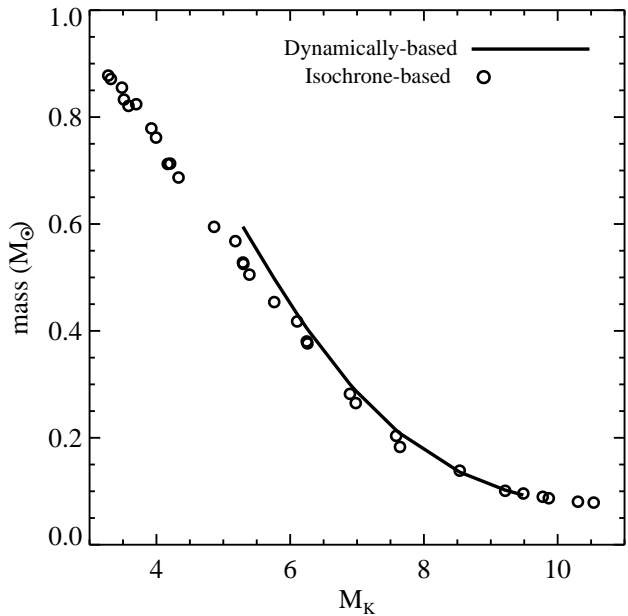}}
\caption{Stellar mass--K-band absolute magnitude relation derived for
  the IRTF stars ({\it symbols}) compared to the relation derived from
  dynamically-based masses.  Dynamical masses are derived from
  analysis of stars in binary orbits \citep{Delfosse00}.  IRTF stellar
  masses are derived from a theoretical $L_{\rm bol}(M)$ relation; see
  the text for details.}
\label{fig:massmk}
\end{figure}

For these reasons $\lbol$ is a very robust parameter estimated for the
IRTF stars.  We therefore use $\lbol$ to estimate the stellar mass for
each IRTF star based on the $\lbol-M$ relation shown in Figure
\ref{fig:iso1} and described in $\S$\ref{s:iso}.  The resulting
mass---K-band luminosity relation for main sequence stars is shown in
Figure \ref{fig:massmk}.  These isochrone-based masses are compared to
the mass-luminosity relation derived for stars in binary orbits by
\citet{Delfosse00}.  The dynamicaly-estimated mass-luminosity relation
is completely independent from our isochrone-based method, and so the
impressive agreement provides strong support to our approach.  Note
that only two stars pin down the dynamically-based $M_K-M$ relation at
$0.45\Msun<M<0.6\Msun$, so the mild divergence between the isochrone
and dynamically-based masses in this range is not a significant source
of concern.  Once a mass is assigned to each star, we can then
estimate its log$(g)$ based on the isochrones.

We have surveyed the literature for independently-derived physical
parameters for our sample of 91 IRTF stars.  We have found parameters
for 35 of the stars (38\% of the sample) in one or more of the MILES,
ELODIE \citep{Prugniel01}, SPOCS \citep{Valenti05}, and Indo-US
\citep{Valdes04, Wu11} libraries.  The mean and standard deviation of
the differences in $\teff$ are $\approx20$ K and $\approx100$ K, while
for log($g$) the mean and standard deviation of the offsets are
$\approx0.05$ and $\approx0.2$.  These differences are comfortably
within the typical errors on $\teff$ and log($g$) and give confidence
to our independently-derived parameters.  The differences are smallest
between our parameters and those derived by the SPOCS survey, which is
encouraging because the SPOCS survey consists of very high resolution
($R\sim70,000$) spectra and very careful consideration of the
available line data in their analysis.  These surveys also provide
estimates of the metallicity of each star.  The average metallicity of
the IRTF stars in common with MILES is $-0.07$, while for stars in
common with SPOCS the average metallicity is $+0.05$.  This supports
our assumption that the IRTF stars are of approximately solar
metallicity.

For two IRTF stars on the subgiant branch (HD10697 and HD25975) we
have adopted the SPOCS and Indo-US values for their $\teff$ and
log($g$) since those values are more accurate (being based on high
resolution spectra), and result in better agreement with the location
of 13.5 Gyr isochrones.  Adoption of these parameters has essentially
no impact on the resulting spectral synthesis, since the $\teff$ and
log($g$) values are only used to pair IRTF and MILES stars.

The MILES library consists of $985$ stars covering a wide range in
atmospheric parameters \citep{Sanchez-Blazquez06}.  The spectra have a
FWHM of 2.5\AA\, and are relative flux calibrated, but not absolute
flux calibrated.  Atmospheric parameters ([Fe/H], log$(g)$, and
$\teff$) were derived for MILES stars in \citet{Cenarro07}.  We use
the MILES library v9.1 \citep{milesv9.1}, which corrects several small
errors in previous releases.  We removed hot variable stars and
binaries (as identified in SIMBAD), and stars whose $\teff$ appeared
to be in error based on visual inspection.  This resulted in 28 stars
being removed from the library.

The two coolest M dwarfs in the MILES library have spectral types M6
and M7.  Both stars are magnetically active, as evidenced by strong
Balmer emission lines in their optical spectra.  Activity occurs in
$\approx60$\% of such stars, and increases to $\approx75$\% for M8V
stars \citep{West04}.  In addition to Balmer emission, these stars
also display strong CaII H \& K lines in emission.  The third coolest
M dwarf, of spectral type M3, also displays weak H \& K emission
lines.  We will return to the importance of magnetic activity and
associated chromospheric emission lines in cool M dwarfs later in the
paper.

\begin{figure}[!t]
\center
\resizebox{3.5in}{!}{\includegraphics{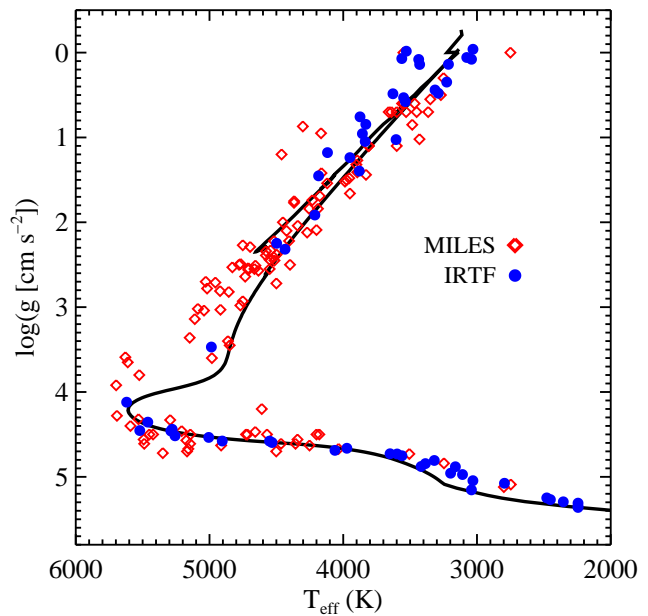}}
\caption{Location in the HR diagram of stars from the empirical IRTF
  and MILES stellar libraries.  Only the subset of IRTF stars used in
  our fiducial model are shown.  Also shown is the fiducial isochrone
  ($Z=Z_\Sol$, $t=13.5$ Gyr).}
\label{fig:imhr}
\end{figure}

The MILES and IRTF libraries are combined in the following way.
First, we select stars from MILES that have a metallicity of
$-0.1<$[Fe/H]$<0.1$.  This leaves 197 stars in the MILES library.  We
then take each IRTF star and pair it to the MILES library via
interpolation in $\teff-{\rm log}(g)$ space.  Figure \ref{fig:imhr}
shows how MILES and IRTF stars populate this space.  The MILES library
is considerably larger than the IRTF library, so this pairing is
straightforward.  Once the pairing is made, the interpolated MILES
spectrum is normalized against the $V-$band magnitude of the IRTF
star.  With this procedure we obtain an empirical spectrum for each
star in $\teff-{\rm log}(g)$ space that spans the wavelength range
$0.35\mu m<\lambda<2.4\mu m$ with only a modest gap at $0.74\mu
m<\lambda<0.81\mu m$.

The astute reader will notice in Figure \ref{fig:imhr} that there are
no MILES dwarf stars cooler than $\teff=2750$ K, corresponding to
$M=0.15\Msun$.  Because of this, the five IRTF stars cooler than this
are paired to the coolest MILES star available.  As MILES spectra are
always normalized to the $V-$band photometry of the IRTF stars, the
overall change in $V-$band luminosity with mass is correctly modeled
with our approach.  It is only the surface gravity and
temperature-sensitive lines that are affected by the lack of cooler
dwarfs in the MILES library.  Finally, we point out that $<0.15\Msun$
stars contribute ten times less flux at $0.5\mu m$ compared to $>1\mu
m$, so the approximate treatment of these stars in the optical should
not be a source of concern.

\subsection{Synthetic spectral libraries}
\label{s:sspec}

Our empirical library, by construction, is very well suited for
generating integrated light spectra of stellar populations with
approximately Solar abundance ratios. However, it does not allow us to
investigate changes in the overall metallicity or changes in the
abundances of individual elements. We turn to synthetic spectra to
incorporate this option in our modeling. These synthetic spectra are
used to calculate {\em relative} changes with respect to a fiducial
synthetic model, and these relative changes are then applied to
indices that we measured from our empirical spectra \citep[as in
e.g.,][]{Trager00, Thomas03, Schiavon07}.  A synthetic spectral
library does not exist that meets our needs, so we have generated our
own library specifically for this project.

\begin{figure}[!t]
\center
\resizebox{3.5in}{!}{\includegraphics{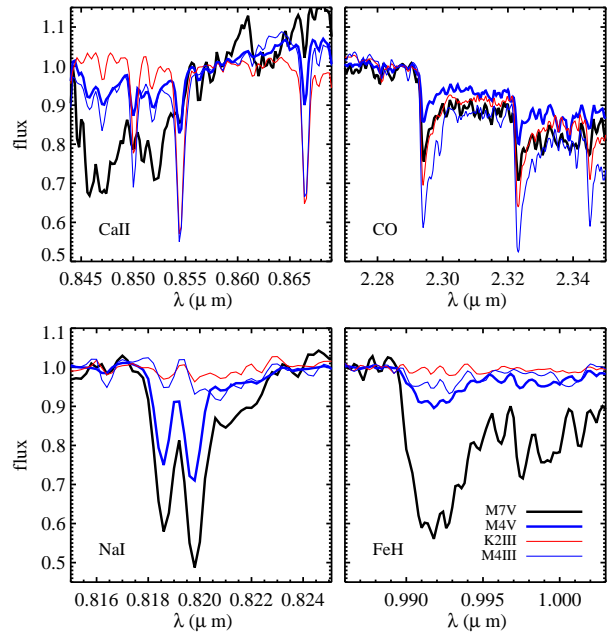}}
\vspace{0.1cm}
\caption{IRTF spectra for stars of various spectral types and
  luminosity classes.  The plots feature four surface
  gravity-sensitive and temperature-sensitive spectral features.  The
  top panels highlight features that are strong in giants compared to
  dwarfs, while the bottom panels highlight features that are strong
  in dwarfs compared to giants.  Spectra are normalized in order to
  highlight the gravity-sensitive lines.  Notice in particular the two
  spectra that have the same spectral type (M4) but different
  luminosity class (V vs. III).}
\label{fig:starspec}
\end{figure}

\begin{deluxetable}{lcccl}
\tablecaption{Definition of Spectral Indices}
\tablehead{ \colhead{Index \tablenotemark{a}} &\colhead{Feature}
  &\colhead{Blue} &\colhead{Red}
  &\colhead{Notes} \\ \colhead{} & \colhead{}&\colhead{
    Continuum}&\colhead{Continuum}&\colhead{} \\ \colhead{} & \colhead{(\AA)}&\colhead{
    (\AA)}&\colhead{(\AA)}&\colhead{}
}
\startdata
CaII0.39   & $3899.5-4003.5$ & $3806.5-3833.8$ & $4020.7-4052.4$ & 1 \\
CN0.41    & $4143.3-4178.3$ & $4085.0-4097.6$ & $4245.5-4285.5$ & 2 \\
CaI0.42   & $4223.5-4236.0$ & $4212.2-4221.0$ & $4242.2-4252.2$ & 2 \\
CH0.43    & $4282.6-4317.6$ & $4267.6-4283.8$ & $4320.1-4336.5$ & 2 \\
FeI0.44   & $4370.3-4421.6$ & $4360.3-4371.6$ & $4444.1-4456.6$ & 2 \\
C$_2$0.47 & $4635.3-4721.6$ & $4612.8-4631.6$ & $4744.1-4757.8$ & 2 \\
FeI0.50   & $4979.2-5055.4$ & $4947.9-4979.2$ & $5055.4-5066.7$ & 2 \\
MgH0.51   & $5070.5-5135.6$ & $4896.5-4959.0$ & $5302.6-5367.6$ & 2\\
FeI0.52   & $5247.2-5287.2$ & $5234.7-5249.7$ & $5287.2-5319.7$ & 2 \\
MgI0.52a  & $5161.6-5194.1$ & $5144.1-5162.8$ & $5192.8-5207.8$ & 2 \\
MgI0.52b  & $5165.0-5220.0$ & $5125.0-5165.0$ & $5220.0-5260.0$ & 3 \\
FeI0.53   & $5313.6-5353.6$ & $5306.1-5317.4$ & $5354.9-5364.9$ & 2 \\
NaI0.59   & $5878.5-5911.0$ & $5862.2-5877.2$ & $5923.7-5949.7$ & 2 \\
NaI0.82   & $8177.0-8205.0$ & $8170.0-8177.0$ & $8205.0-8215.0$ & 3 \\
CaII0.86  & $8484.0-8513.0$ & $8474.0-8484.0$ & $8563.0-8577.0$ & 4,5 \\
          & $8522.0-8562.0$ & $8474.0-8484.0$ & $8563.0-8577.0$ &   \\
          & $8642.0-8682.0$ & $8619.0-8642.0$ & $8700.0-8725.0$ &   \\
MgI0.88   & $8801.9-8816.9$ & $8777.4-8789.4$ & $8847.4-8857.4$ & 6 \\
TiO0.89   &                 & $8835.0-8855.0$ & $8870.0-8890.0$ & 3,7\\
FeH0.99   & $9905.0-9935.0$ & $9855.0-9880.0$ & $9940.0-9970.0$ & 3 \\
NaI1.14   & $11372-11415$ & $11360-11370$ & $11417-11427$ & 3 \\
KI1.17    & $11680-11705$ & $11667-11680$ & $11710-11750$ & 3,5 \\
          & $11765-11793$ & $11710-11750$ & $11793-11810$ & \\
AlI1.31   & $13115-13165$ & $13090-13113$ & $13165-13175$ & 3 \\
CaI1.98   & $19770-19795$ & $19740-19765$ & $19800-19840$ & 3,5 \\
          & $19845-19880$ & $19800-19840$ & $19885-19895$ &   \\
NaI2.21   & $22047-22105$ & $22035-22045$ & $22107-22120$ & 3 \\
CO2.30    & $22932-22982$ & $22860-22910$ & $23020-23070$ & 3,5 \\
          & $23220-23270$ & $23150-23200$ & $23300-23350$ &   \\
\enddata
\vspace{0.1cm} 
\tablecomments{(1) \citet{Serven05}; (2) Lick-based index
  \citep{Trager98}; (3) new index; (4) \citet{Cenarro01}; (5) Index
  with multiple features.  In such cases the total EW is the sum of
  the individual EWs; (6) \citet{Diaz89}; (7) Index defined as a flux
  ratio between blue and red pseudocontinua.}
\tablenotetext{a}{Index names indicate the primary species and central
wavelength in $\mu m$.  Wavelengths are in vacuum.}
\label{t:indx}
\end{deluxetable}

We have made use of the ATLAS model atmosphere and spectrum synthesis
package \citep{Kurucz70, Kurucz93}, ported to Linux by
\citet{Sbordone04}.  In particular, we used the ATLAS12 model
atmosphere code, which utilizes the opacity sampling technique.  The
use of ATLAS12 for our purpose is relatively novel because we allow
the opacity to vary with each elemental abundance variation.  In
previous work related to the response of spectral indices to
individual abundance variations \citep[e.g.,][]{Tripicco95, Korn05}, a
single opacity distribution function was used for all individual
abundance variations.  The atmospheres are plane-parallel and
stationary.  The standard mixing-length theory is adopted for the
treatment of convection, and local thermodynamic equilibrium (LTE) is
assumed.

Once the atmospheres are generated, the SYNTHE program
\citep{Kurucz81} is used to generate a very high resolution
($R=100,000$) spectrum for each star over the wavelength range $0.3\mu
m<\lambda<3.0\mu m$.  A microturbulent velocity of $V_t=2.0\kms$ is
adopted, independent of spectral type and luminosity.  We have
recomputed a subset of the models with $V_t=1\kms$ in order to explore
the sensitivity of the spectra to this parameter.  This change in
$V_t$ results in a sizable change in the equivalent widths of many
features \citep[$\approx0.2-2$\AA; see also][]{Tripicco95}.  We care
only about the relative response of the models (see below), and this
is much less effected by a change in $V_t$. For example, considering
the response of the model spectra to a factor of two enhancement in
iron abundance, the difference between $V_t=2\kms$ and $1\kms$ is less
than 0.1\AA\, in equivalent width for all lines of interest.

The line data are critical for construction of realistic stellar
atmospheres and spectra.  We have adopted the atomic and molecular
linelists available on Kurucz's
webpage\footnote{\texttt{http://kurucz.harvard.edu/}}, including the
molecules H$_2$O, TiO, FeH, C$_2$, CH, CN, NH, SiO, SiH, OH, MgH, CO,
and H$_2$.  In particular, we use the TiO linelist from
\citet{Schwenke98} and the H$_2$O linelist from \citet{Partridge97}.
The FeH linelist is from the work of \citet{Dulick03}, but with a
partition function tabulated by Kurucz specifically for this project.

We have identified 20 points along the 13.5 Gyr isochrone shown in
Figure \ref{fig:iso1} ranging from the tip of the RGB down to
$\teff=3170$ K and log$(g)=5.05$, corresponding to a main sequence
mass of $0.15\Msun$.  A synthetic spectrum is computed at each of
these points for later use in the spectral synthesis.  Model
atmospheres for cooler dwarfs failed to converge.  We therefore
extrapolate our spectral library in $\teff$ to cooler dwarfs down to
0.08$\Msun$.  This approximation will have little impact on our
results because we only use the model spectra to created integrated
light spectra for a Chabrier IMF, for which stars with
$0.08\Msun<M<0.15\Msun$ contribute $\approx10^{-3}$ of the total flux.
Notice that previous work \citep[e.g.,][]{Tripicco95, Korn05,
  Serven05} only considered two or three points along the isochrone in
order to gauge the relative response of spectral features to changes
in abundances.

In the present work we adopt the latest solar abundances from
\citet{Asplund09}.  These new abundances for C, N, O, and Ne are
significantly lower than the old solar abundances of \citet{Anders89}
and \citet{Grevesse98}.  In particular, the new [O/H] abundance is
0.24 dex lower than the \citet{Anders89} abundances.  This results in
substantially better agreement between model spectra and data for
late-type M dwarfs due to the resulting weaker H$_2$O opacity.  

We have run models with individual abundance variations for C, N, Na,
Mg, Si, Ca, Ti, Fe, and generic $\alpha-$elements\footnote{We include
  the elements O, Ne, Mg, Si, S, Ca, and Ti as ``$\alpha$'' elements
  in the present work.}, although in the present work we focus
primarily on the elements C, Na, Ca, Fe, and $\alpha$.  We have also
run models with a factor of two variation in the total metallicity,
$Z$, with abundance ratios fixed at their solar values.  We vary each
abundance both higher {\it and lower} than its canonical value.  This
is in contrast to previous work, which only increased the abundances
of elements in order to gauge their effect on the spectrum.

\subsection{Spectral indices}
\label{s:indices}

\begin{figure*}[!t]
\center
\resizebox{6.5in}{!}{\includegraphics{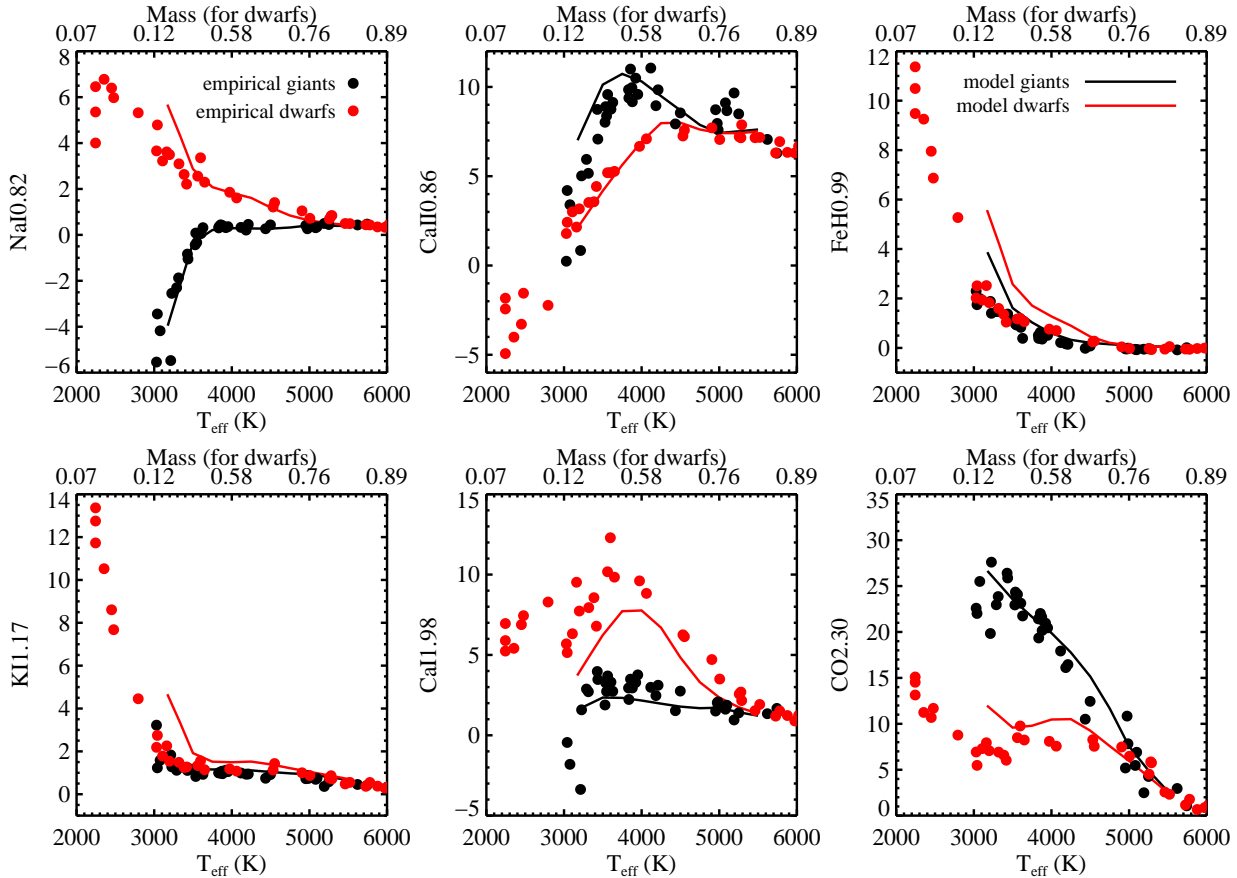}}
\vspace{0.1cm}
\caption{Dependence of selected spectral indices on effective
  temperature for dwarfs ({\it red points and lines}) and giants ({\it
    black points and lines}).  The empirical IRTF spectra ({\it
    symbols}) are compared to the model synthetic spectra ({\it
    lines}).  At the top of each panel $\teff$ is converted into
  stellar mass for dwarfs via the isochrones discussed in
  $\S$\ref{s:iso}.}
\label{fig:indteff}
\end{figure*}

In later sections we will rely on spectral indices in order to
investigate the sensitivity of various spectral features to the
parameters age, metallicity, and IMF.  Spectral indices have been
utilized for decades to compress the amount of information in a
spectrum.  The most well-known are the Lick/IDS indices
\citep{Burstein84, Worthey94b, Trager98}.  These indices reside in the
blue region of the spectrum ($4000$\AA$<\lambda<6400$\AA), and include
indices sensitive to CN, C$_2$, CH, Ca, Cr, Fe, Mg, MgH, Na, and TiO
\citep{Tripicco95, Korn05}.  In the following work, we will utilize
the Lick index wavelength intervals for the index bandpass and
pseudocontinua as defined in \citet{Trager98}, but we make no effort
to place our indices on the zero-point system of the Lick
system. These indices are defined in Table 1.  All indices in the
present work (except for TiO0.89) are quoted as equivalent widths
(EWs) with units of \AA\, \citep[the procedure for measuring EWs is
described in][]{Worthey94b}.  All wavelengths in this paper are quoted
in vacuum.

We supplement the Lick-based blue indices with additional indices as
defined in Table 1.  These indices were chosen for their promise in
constraining the number of low mass stars from integrated light (in
practice, we selected features that changed by more than 1\% between a
Chabrier and $x=3$ IMF; see Figure \ref{fig:specrat} below).  The
majority of the new indices are defined and measured in a manner
analogous to the Lick-based indices.  Exceptions include the calcium
triplet (CaT) at $0.86\mu m$, which is comprised of three distinct
features and so the total EW is the sum of the EWs of the individual
features \citep[see][for details]{Cenarro01}.  Likewise, the CaI
feature at $1.98\mu m$ is a doublet even when doppler broadened to
$\sigma=300\kms$, and so this feature is defined as the sum of two
EWs.  The CO index at $2.3\mu m$ is defined by the first two bandheads
of the $^{12}$CO molecule.  Finally, the TiO0.89 index is the only
index defined as a ratio of fluxes between blue and red
pseudocontinua, rather than an equivalent width.

The names given to the indices in Table 1 highlight the primary
species controlling the strength of each index.  In reality, each
index is sensitive to a number of atomic and molecular species
\citep[see e.g.,][for details]{Tripicco95, Korn05}.

The use of indices suffers from several well-known problems.  The EW
of an absorption feature is measured with respect to a
pseudocontinuum, which itself is composed of absorption features.  The
variation of an index due to changes in metallicity or age may
therefore be due to a combination of changes in the feature itself and
changes in the pseudocontinuum.  Moreover, each index is in reality a
blend of several absorption lines arising from more than one element.
The spectral shape of the feature often contains additional
information, beyond its EW, but this information is lost in an index.
We will highlight these caveats in our results where applicable.  

We emphasize that our use of indices in this work is restricted to
``intuition-building'' in the form of index-index plots --- we do not
advocate using indices when actually fitting data because of the
limitations of indices highlighted above.  The model spectra will be
made available upon request so that others can measure their own
indices and/or directly compare models to data at the appropriate
resolution.

\subsection{Behavior of the stellar libraries}
\label{s:bsl}

Figure \ref{fig:starspec} shows surface gravity-sensitive spectral
regions for selected IRTF stars.  This figure highlights the potential
power of certain spectral features at discriminating between dwarfs
and giants.  In this figure an M4 dwarf is compared to an M4 giant.
These two stars have the same spectral type and thus broadly the same
SED shape, but clearly have very different strengths of these spectral
features. The M4 giant displays no FeH lines but instead shows TiO
lines at almost the same wavelength as FeH.  Also notice the dramatic
increase in the FeH strength toward the latest M dwarfs.  These stars
also show very different features around the CaII lines compared to
both slightly warmer dwarfs and giants.  The different relative
response of the NaI and FeH lines to later-type M dwarfs implies that
these lines are sensitive to IMF variations in different ways.

In Figure \ref{fig:indteff} we show selected spectral indices as a
function of effective temperature for giants and dwarfs.  This figure
compares the empirical IRTF spectra to the synthetic spectra.  As
expected, the indices NaI0.82 and FeH0.99 are stronger in dwarfs than
in giants, and FeH0.99 becomes very strong for the coolest dwarfs.
The CO2.30 and CaII0.86 indices also behave as expected in that they
are stronger in giants than dwarfs.  Notice though that the CO2.30
index becomes progressively stronger for dwarfs with $\teff<3000$ K
(see Figure \ref{fig:starspec}).

Two new spectral regions are explored in this figure: the KI doublet
at 1.17$\mu m$ and the CaI doublet at 1.98$\mu m$.  These features are
both stronger in dwarfs than in giants, but they behave in very
different ways.  The KI1.17 index is even more sensitive to the very
coolest stars than FeH0.99, while CaI1.98 peaks in strength at
$\approx3700$ K (corresponding to $\sim0.55\Msun$ for dwarfs).

It is noteworthy that the synthetic spectra reproduce the qualitative
behavior of all the indices shown in Figure \ref{fig:indteff} and
reproduce {\it in detail} the NaI0.82, CaII0.86, KI1.17, and CO2.30
indices.  The models extend only to 3170K, so the models cannot be
compared to the coolest M dwars.  The synthetic spectra predict a
CaII0.86 index stronger than observed for the coolest giants, which
might be related to non-LTE effects, a FeH0.99 index stronger than
observed for most dwarfs, which is probably due to problems with the
FeH linelist and/or partition function, and a CaI1.98 index for dwarfs
that is weaker than observed, which may be due to the fact that the
pseudocontinuum is strongly affected by H$_2$O in this wavelength
region.

The trends in this figure set the stage for the results to be
discussed in $\S$\ref{s:res} and in particular $\S$\ref{s:imfi} where
we discuss the ability of combinations of these indices to constrain
the {\it shape} of the low-mass IMF in the integrated light of old
stellar populations.

\begin{figure*}[!t]
\center
\resizebox{7in}{!}{\includegraphics{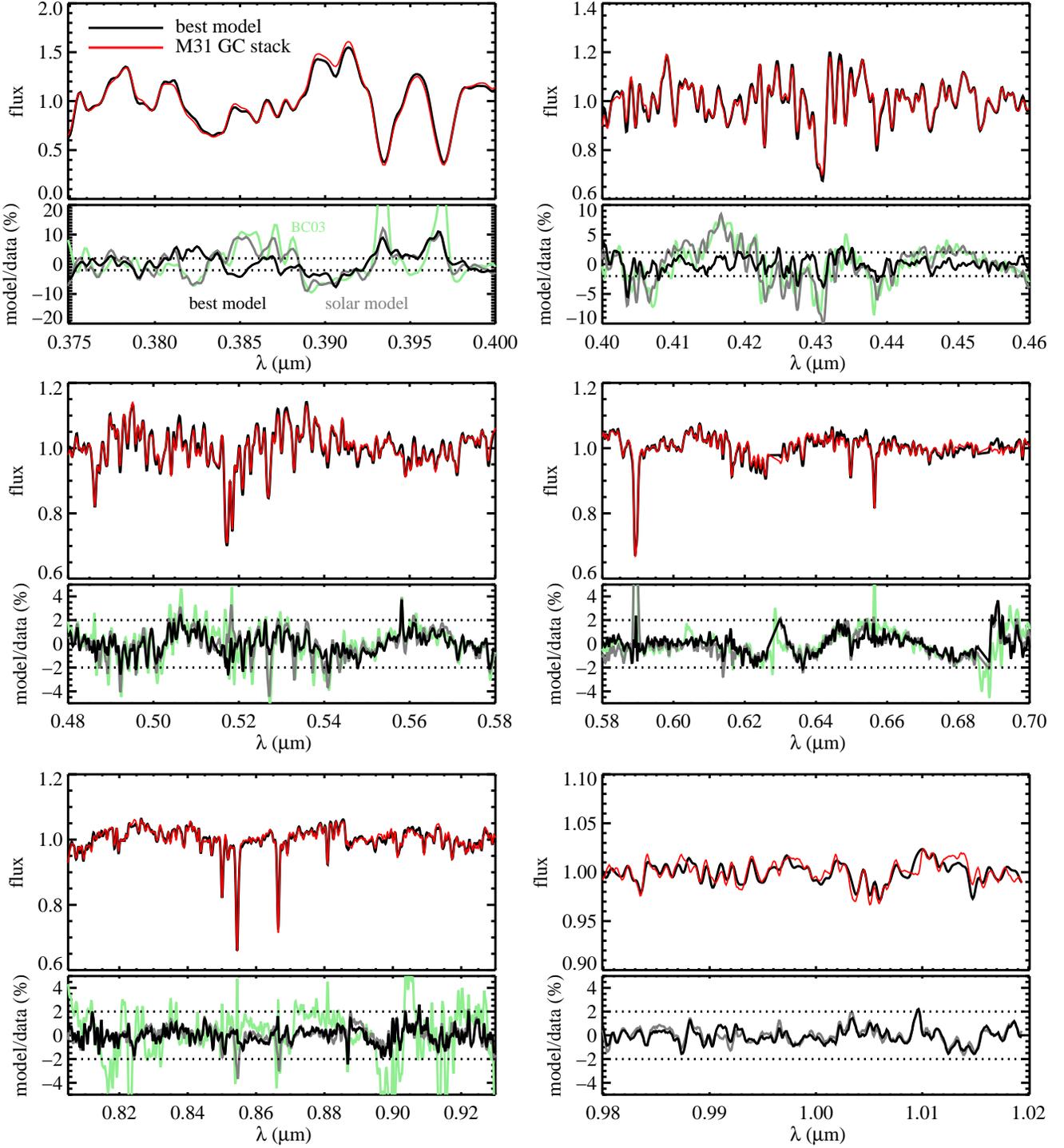}}
\vspace{0.1cm}
\caption{Comparison of a stacked spectrum of four metal-rich M31 GCs
  to models.  A best-fit model with variable abundance ratios (the
  ``best model''), is compared to our solar metallicity model and the
  high-resolution solar metallicity model from
  \citet[][BC03]{Bruzual03}.  Both model and data spectra were
  continuum normalized by a third-order polynomial within each
  wavelength interval.  The bottom portion of each panel shows the
  ratio between model and data in percent units.  The dotted lines
  demarcate a $\pm2$\% deviation.  For the best-fit model, the
  deviations are less than 2\% redward of $0.4\mu m$ and the
  root-mean-square deviation is 1.2\%, testifying to the high fidelity
  of our models.}
\label{fig:gcspec}
\end{figure*}

\subsection{Synthesis}
\label{s:syn}

With the isochrones and stellar libraries described in previous
sections we are now able to construct integrated light SSP spectra via
Equation 1.  We emphasize that the synthesis includes not only the RGB
but also the HB and AGB.  For a 13.5 Gyr solar metallicity population
with a Chabrier IMF, stars on the HB and AGB comprise $\approx25$\% of
the integrated light at $1\mu m$, so their inclusion in the synthesis
is essential.  When adding the HB and AGB we frequently used an
empirical star that was already counted on the RGB\footnote{While this
  is common practice, some authors use only genuine AGB stars in the
  synthesis \citep[e.g.,][]{Marmol08}.}.  In such cases only the
stellar mass is changed in accordance with the $\lbol-M$ relation on
the HB and AGB.

Abundance effects are implemented in a differential sense.  This is
the standard method of implementing abundance ratio effects in
population synthesis \citep[e.g.,][]{Trager00, Thomas03, Schiavon07,
  Walcher09}.  The empirical stellar libraries constitute the base set
of models, and the synthetic stellar libraries, which exist for
arbitrary abundance patterns, are used to differentially change the
base models.  In practice, the model makes a prediction for an SSP
spectrum of arbitrary abundance pattern $R$ via:
\noindent
\be
f(R) = f_{\rm base} \frac{f_{\rm syn}(R)}{f_{\rm syn}(Z_{\Sol})},
\ee
\noindent
where $f$ is the spectrum and we have implicitly assumed that the base
empirical models have solar metallicity.  Indices are then measured
from these model spectra.

Our model thus provides SSPs for arbitrary IMFs, for stellar ages in
the range $3-13.5$ Gyr, and for arbitrary elemental abundance
patterns.  We have avoided younger ages both because the IRTF library
does not contain hot stars and because at younger ages
thermally-pulsating AGB stars, whose evolution is highly uncertain,
contribute significantly to the near-IR flux.

We have performed a number of tests to ensure that our resulting base
model is not affected by one or more `pathological' empirical stellar
spectra.  In addition to investigating each empirical spectrum by eye,
we have constructed a set of models by removing one stellar spectrum
at a time from the library and constructing a new SSP based on these
$N-1$ spectra.  The resulting indices never changed by more than 2\%
from our fiducial model.  As will be shown in later sections, the
index variations induced by modest variations in age, IMF, and
abundances are always much larger than 2\%.  We conclude that our
results are not unduly sensitive to any particular star in the
empirical library.

\section{A Brief Comparison to Data}
\label{s:data}

In this section we briefly explore the quality of the models
constructed in the previous section by comparing them to spectra of
globular clusters (GCs).  There are a number of massive metal-rich
([Fe/H]$\gtrsim-0.1$) ancient GCs in M31 \citep{Caldwell11}.  Such
objects are ideal to compare to our models because our models are
applicable to old metal-rich systems and because massive GCs will have
a fully populated RGB.  Red and near-IR spectra were obtained for the
GCs B143, B147, B163, and B193 from the LRIS spectrograph on the Keck I
telescope.  These data were described in \citet{vanDokkum11}.  Blue
spectra were obtained from the Hectospec spectrograph on the MMT by
Nelson Caldwell (unpublished).  The spectra were stacked to create an
average GC spectrum with typical signal-to-noise $>100$.  There are
significant sky subtraction issues in the wavelength regions $0.7\mu
m<\lambda<0.8\mu m$ and $0.93\mu m<\lambda<0.98\mu m$ so those regions
will be omitted in the following comparison.

Figure \ref{fig:gcspec} shows the comparison between our models and
the stacked M31 GC spectrum.  Within each wavelength interval the
spectra have been continuum normalized by a third-order polynomial.
Our best-fit model was obtained via a Markov Chain Monte Carlo fitting
technique with variation in each of the elements C, N, Na, Mg, Si, Ca,
Ti, Fe, and with lock-step variation in the elements O, S, and Ne.
The IMF is also varied, and the age is kept fixed at 13.5 Gyr.  The
data provide very strong constraints on the abundances of these
elements and set a strong upper bound on the slope of the IMF.
Discussion of the derived parameters is beyond the scope of this paper
but is the focus of ongoing work.

The purpose of Figure \ref{fig:gcspec} is to demonstrate the high
quality of our models.  The rms difference between the data and our
best-fit model is 1\%, and the percent difference at any wavelength
point rarely exceeds 2\% at $\lambda>0.4\mu m$.  In this figure we
also include comparison to our solar metallicity model, and the
high-resolution solar metallicity model from \citet{Bruzual03}.  Both
of these models assume a Chabrier IMF.  The solar metallicity models
provide a reasonable fit to the data over much of the optical
wavelength range, but at essentially every wavelength the best-fit
model performs better.  Regions where the best-fit model substantially
outperforms the solar models include the CH and CN bands in the blue,
the Mg{\it b} doublet at 0.52$\mu m$, the FeI lines at $0.48-0.54\mu m$,
the NaD line at $0.59\mu m$, and the CaII triplet at $\sim0.86\mu m$.

\begin{figure}[!t]
\center
\resizebox{3.5in}{!}{\includegraphics{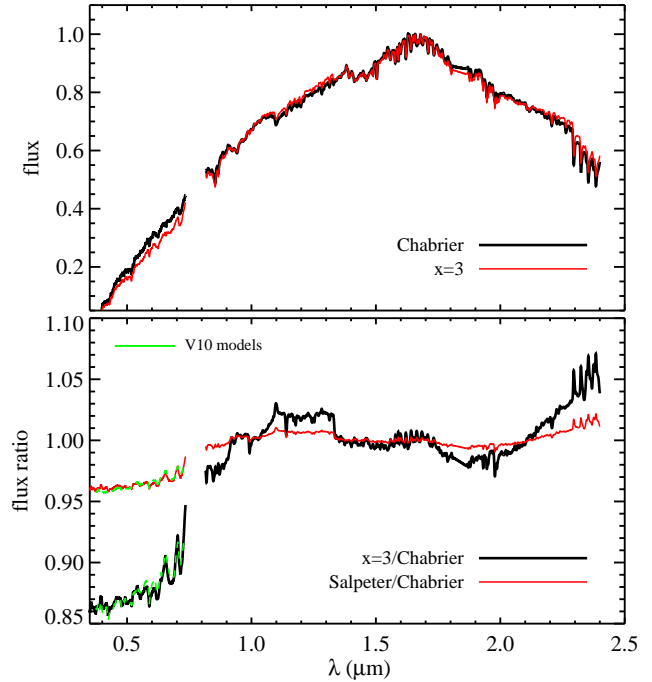}}
\caption{Model spectra for solar metallicity and an age of 13.5 Gyr.
  Spectra are shown for three IMFs and have been smoothed by a
  Gaussian with FWHM$=40$\AA\, in order to highlight the broadband
  differences between models with different IMFs.  The gap at
  $0.74-0.81\mu m$ is due to the gap in wavelength coverage between
  the empirical MILES and IRTF spectral libraries.  {\it Top panel:}
  Comparison between models with a Chabrier and an $x=3$ IMF.  Spectra
  are normalized to unity at 1.6$\mu m$.  {\it Bottom panel:} Flux
  ratio between models with different IMFs, as indicated in the
  legend.  Also shown are ratios of SSPs with different IMFs from
  \citet[][{\it green lines}]{Vazdekis10}.}
\label{fig:over}
\end{figure}

\begin{figure*}[!t]
\center
\resizebox{3.3in}{!}{\includegraphics{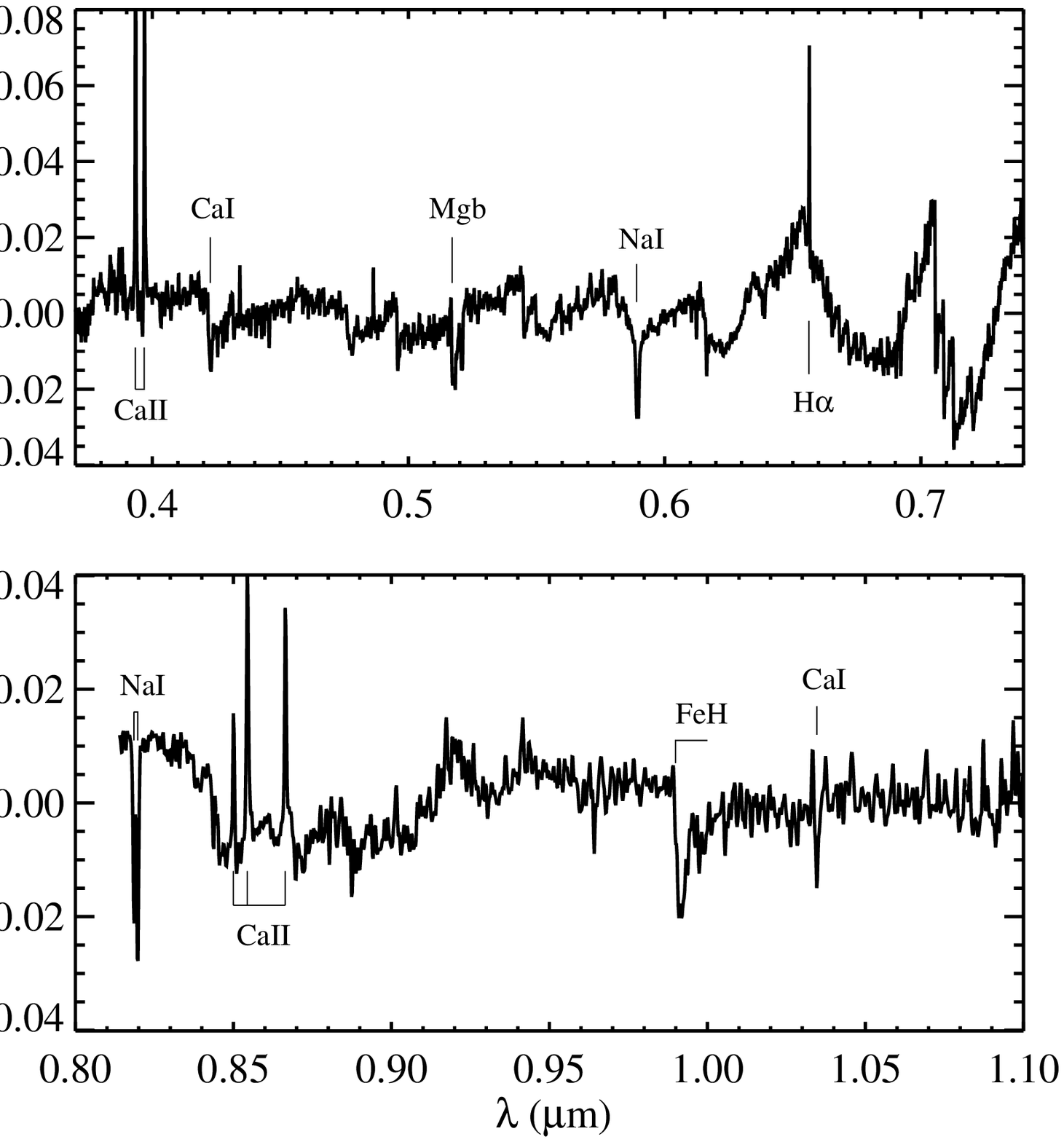}}
\resizebox{3.3in}{!}{\includegraphics{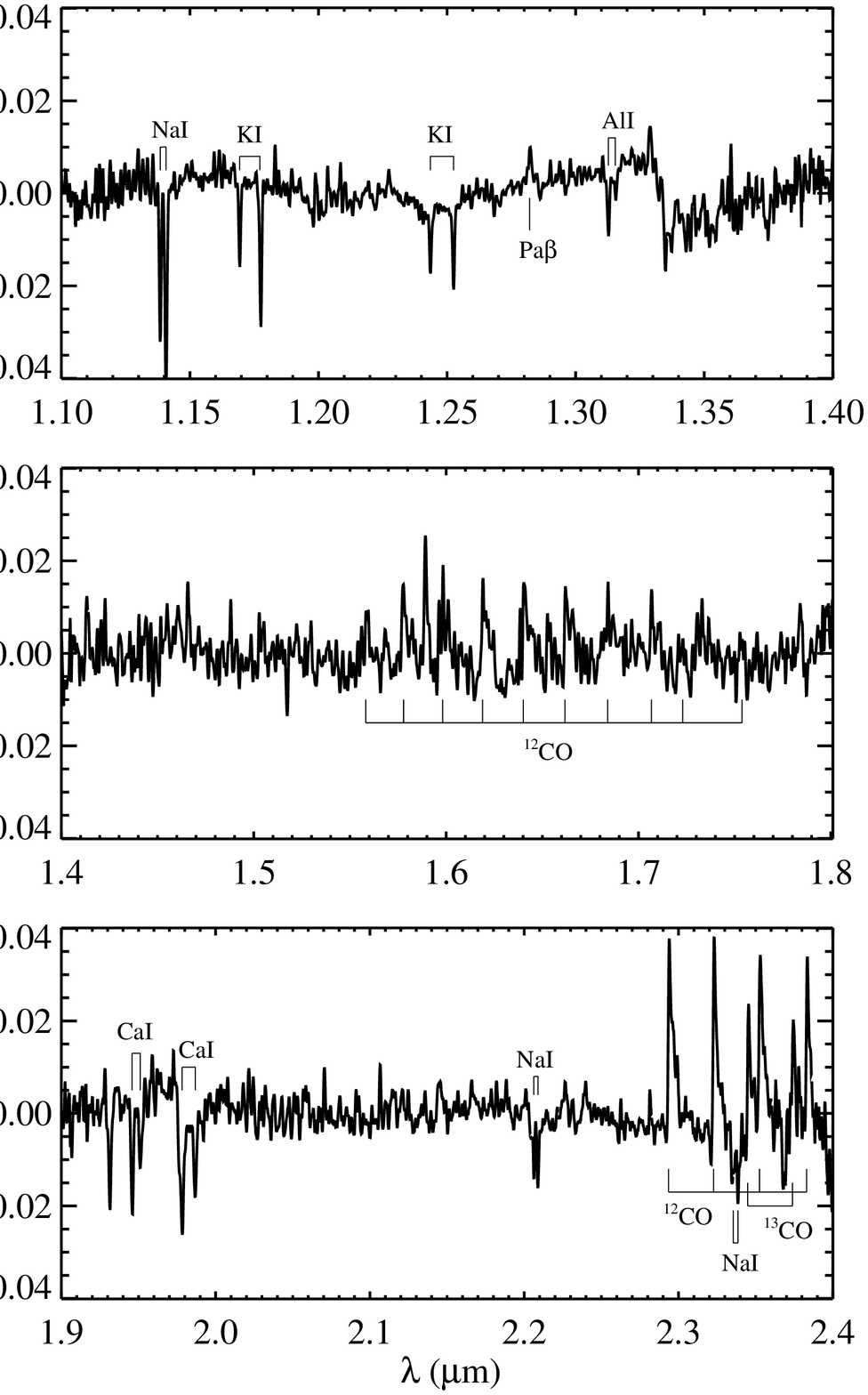}}
\caption{Ratio between a model spectrum with a Chabrier IMF and one
  with an $x=3$ IMF.  The spectra are otherwise identical, both with
  an age of 13.5 Gyr and solar metallicity.  The spectra are at the
  native resolution of the empirical libraries used to generate the
  models ($R\approx2000$), and have been divided by a fourth-order
  polynomial in order to focus attention on the narrow-band features.
  Selected spectral features are labeled.  Notice the different
  $y-$axis range in the upper left panel.}
\label{fig:specrat}
\end{figure*}

This figure demonstrates that the models we have constructed are
capable of reproducing observed spectra at the percent level over the
entire wavelength range of $0.37\mu m<\lambda<1.02\mu m$.  This
comparison also places strong limits on the presence of telluric lines
in our model spectra because the M31 GC spectra are at different
observed wavelengths than our empirical stellar library (The recession
velocity of M31 is $-300\kms$, which means that the GC spectra were
shifted on average $5-10$\AA\, between the observed and restframe).
Evidently, sky subtraction and sky absorption corrections have not
introduced errors at the level $\gtrsim$1\% in either the synthesized
spectra or the M31 spectra.

\begin{figure}[!t]
\center
\resizebox{3.5in}{!}{\includegraphics{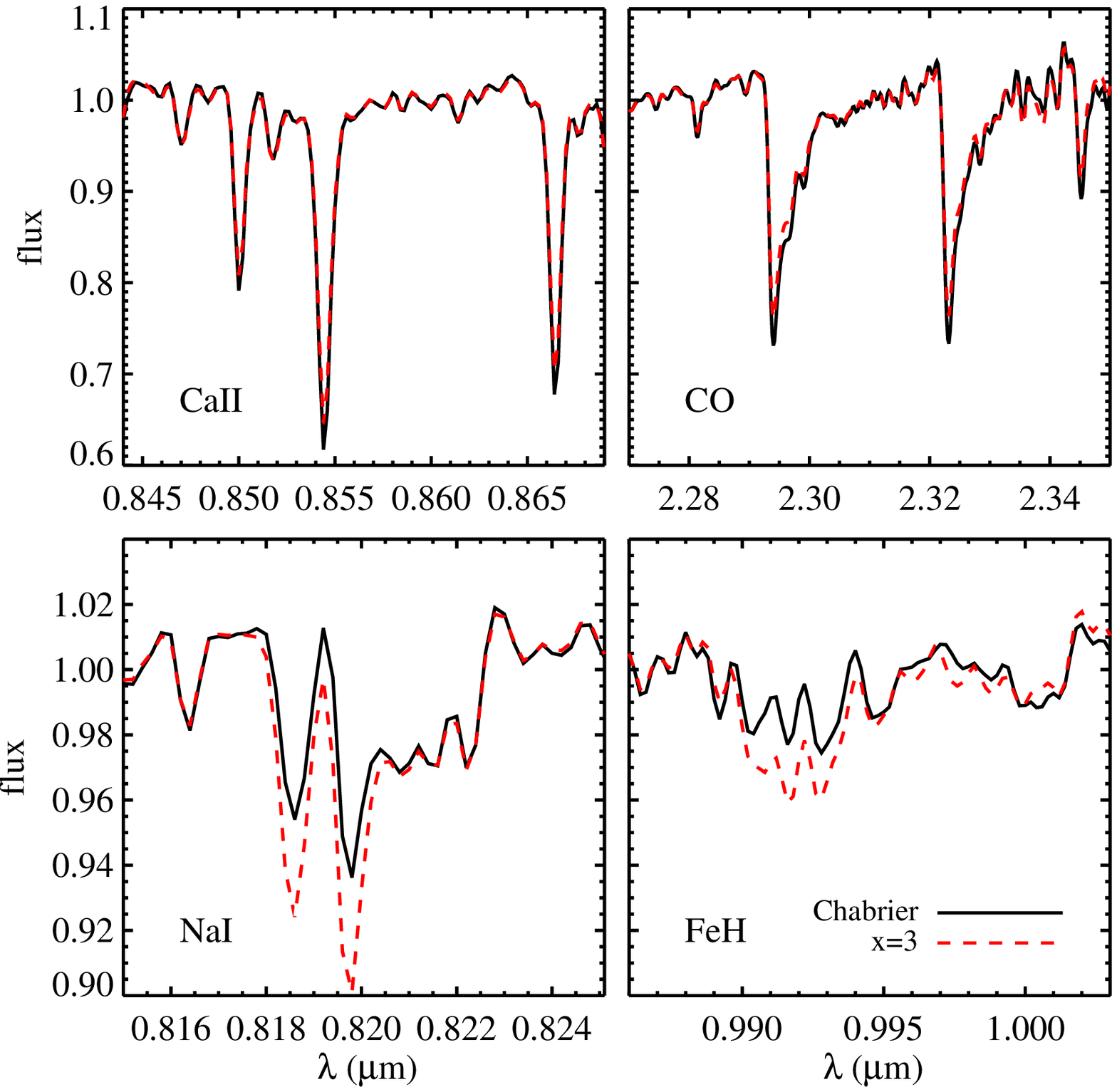}}
\caption{Model spectra for four IMF-sensitive spectral features.
  Models are shown for an age of 13.5 Gyr, solar metallicity, and two
  IMFs: a Chabrier IMF and a bottom-heavy $x=3$ IMF.  The dwarf-strong
  features are shown in the bottom panels, while the giant-strong
  features are shown in the top panels.  Spectra have been divided
  through by a first-order polynomial (excluding the features of
  interest from the fit) in order to highlight the differences in the
  features.}
\label{fig:zoom}
\end{figure}

\section{Behavior of the Models}
\label{s:res}

In this section we discuss how the models behave as a function of the
IMF, stellar age, and elemental abundances.  The goal of this section
is to demonstrate that the integrated light spectrum of an old
population is sufficient to jointly constrain the mean stellar age,
metallicity, abundance pattern, and low mass IMF ($0.08\Msun<
M<1\Msun$).  All of the models presented in this section are SSPs ---
that is, they are coeval and mono-metallic.  Throughout this section
we frequently highlight the IMF-dependence of the integrated spectra
by contrasting a Chabrier IMF with a relatively extreme bottom-heavy
$x=3$ IMF.  Also, in this section we rely on spectral indices in order
to explore the behavior of the models.  However, as discussed in
$\S$\ref{s:indices}, we advocate using the spectra directly, rather
than constructing indices, when fitting models to data.

\subsection{Overview}

Our models based on empirical stellar spectra are shown in Figures
\ref{fig:over}, \ref{fig:specrat}, and \ref{fig:zoom}.  In this series
of figures we explore the sensitivity of the SSP spectrum to the IMF,
zooming in on smaller wavelength intervals with each successive
figure.  The age is fixed at 13.5 Gyr and the metallicity is solar.

In the top panel of Figure \ref{fig:over} we show the overall spectral
energy distribution for two models that differ only in the adopted
IMF.  The bottom panel shows the ratio between spectra constructed
with different IMFs.  In both panels the spectra are normalized to
unity at 1.6$\mu m$.  The broad-band differences between a Chabrier
IMF and a Salpeter IMF in the shape of the spectrum are slight, less
than a few percent \citep[see also][]{Conroy09a}.  The differences
become more substantial for a bottom-heavy IMF with $x=3$.  In this
case $I-z$ colors can differ by $>10$\% in flux.  We caution however
that tying the IRTF spectra (which comprise the model predictions at
$\lambda>0.81\mu m$) to the MILES spectra (which comprise the model
predictions at $<\lambda<0.74\mu m$) relies on a heterogeneous
compilation of $V-K$ colors that are probably not accurate to better
than a few percent.  The absolute broad-band model predictions should
therefore be treated with some caution until the relative flux
calibration between the MILES and IRTF libraries can be independently
assessed.

In this figure we also include predictions from the models of
\citet[][V10]{Vazdekis10}.  These models adopt the MILES stellar
library and the Padova isochrones in constructing SSPs.  The models
shown are of solar metallicity and have an age of 12.6 Gyr.  For the
V10 models we plot ratios between a Salpeter and \citet{Kroupa01} IMF,
and between an $x=2.8$ and Kroupa IMF.  The Kroupa IMF is essentially
identical to the Chabrier IMF, and the $x=2.8$ IMF was the closest in
their model grid to our $x=3$ IMF.  Overall the agreement between our
models and the V10 models is excellent.  On a more refined level the
agreement is even more impressive: after dividing the spectra by a
second order polynomial and broadening by 5\AA\, (rather than the
40\AA\, shown in Figure \ref{fig:over}), the models agree in the
predicted spectral variation between a Salpeter and Chabrier IMF to a
level of $\lesssim0.2$\%.

In Figure \ref{fig:specrat} we zoom in on the narrow spectral features
by dividing the spectra in each wavelength interval by a fourth-order
polynomial.  The panels highlight the spectral response to a change in
IMF when all other variables (age, metallicity) are held fixed.
Selected spectral features are labeled.  This figure summarizes all of
the known surface gravity-sensitive lines in cool stars between
$0.4-2.4\mu m$.  The NaI082, CaII086, FeH099, and CO230 features
are clearly visible.  Additional, less well-known lines are also clearly
visible, including NaI, KI, AlI, and CaI, all of which are stronger in
dwarfs than giants.  The numerous lines from CO in the $H-$band are
also visible.  

As noted in $\S$\ref{s:espec}, the faintest M dwarfs in our empirical
library display strong chromospheric emission lines, particularly the
hydrogen Balmer series and the CaII H \& K lines at $0.39\mu m$.  In
the upper left panel of Figure \ref{fig:specrat} we see that the
H$\alpha$ and CaII lines display strong IMF sensitivity.
Approximately one half of the effect is due to chromospheric emission
in the latest M dwarfs in our sample.  In fact, we expect the true
effect to be even larger, since our sample of optical M dwarf spectra
only extends to spectral type M7.  The remaining one half of the IMF
effect in these lines is simply due to the fact that these features in
absorption are weaker in M dwarfs compared to other stars.

Figure \ref{fig:zoom} zooms in further on four classic IMF-sensitive
spectral features.  The spectra have been divided by a first-order
polynomial to focus attention on the features.  It is evident from
this figure that the differences in flux between a Chabrier and an
$x=3$ IMF in these features are of order several percent.  Differences
between a Chabrier and Salpeter IMF (not shown) are of order one
percent.  This figure demonstrates that very high $S/N$ spectra are
required in order to differentiate between IMFs based on these
features.

In this section we have summarized the basic trends evident in our
models based on empirical stellar spectra.  These models confirm a
wide array of previous work that the low-mass IMF, or phrased another
way, the ratio of M giants to M dwarfs, can be inferred directly from
integrated light spectra of old populations.  The fundamental
uncertainty in any modeling approach is the interplay between the IMF,
average stellar age, and abundance pattern.  In the following section
we address this issue in detail.

\begin{figure*}[!t]
\center
\resizebox{3.5in}{!}{\includegraphics{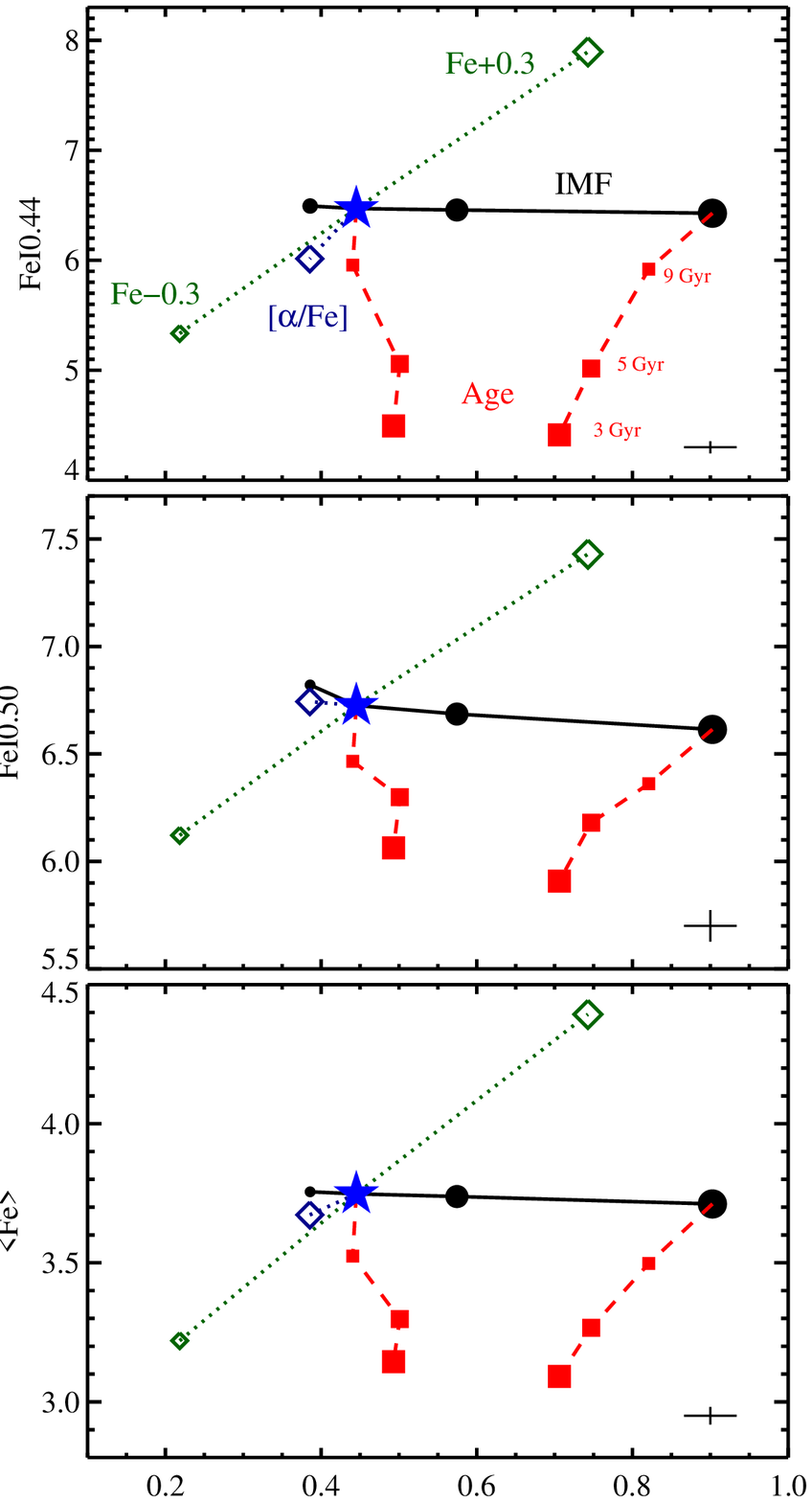}}
\resizebox{3.5in}{!}{\includegraphics{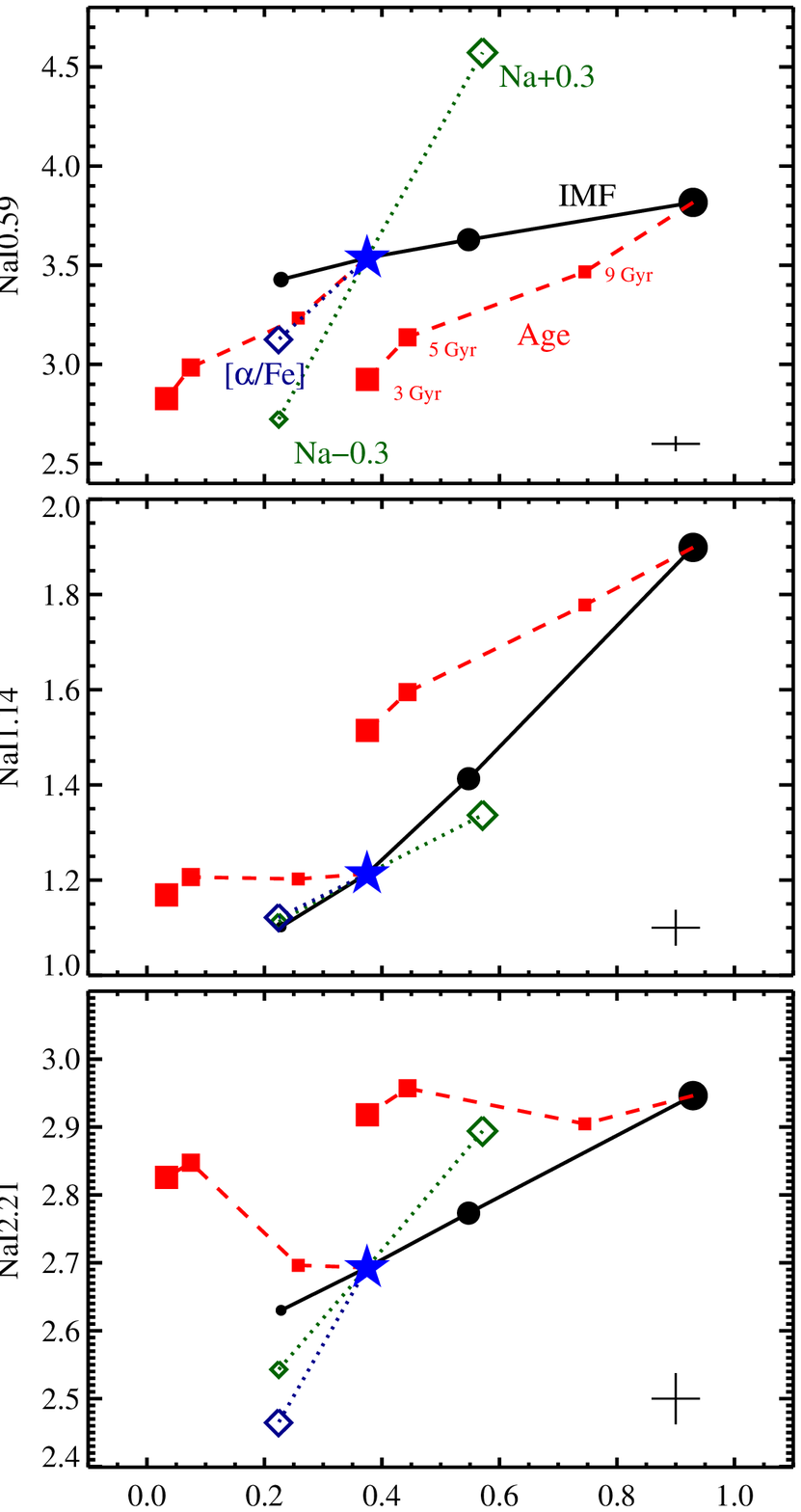}}
\vspace{0.5cm}
\caption{Variation in the EW of spectral indices involving the
  elements iron ({\it left panel}) and sodium ({\it right panel}).
  The panels show the effect of varying the IMF (bottom-light,
  Chabrier, Salpeter, $x=3$; {\it solid lines}), elemental abundance
  ([Fe/H] or [Na/Fe], and [$\alpha$/Fe]; {\it dotted lines}) and
  stellar population age (3,5,9,13.5 Gyr; {\it dashed lines}).  The
  model iron and sodium abundances vary by $\pm0.3$ dex and the model
  $\alpha-$enhancement varies from $0.0-0.2$ dex.  Changes in age are
  only shown with respect to the Chabrier and $x=3$ IMFs for clarity.
  The symbol size increases toward steeper IMFs, higher metallicity,
  and younger ages.  The blue star represents the fiducial model of a
  13.5 Gyr solar metallicity population with a Chabrier IMF.  The EWs
  are in \AA.  Typical measurement uncertainty of each index is shown
  in the lower portion of each panel, assuming a signal-to-noise of
  $S/N=200$ \AA$^{-1}$ and $R\approx2000$.}
\vspace{0.2cm}
\label{fig:FeNa}
\end{figure*}

\subsection{Variation in IMF, age, and elemental abundances}

In this section we combine our model predictions based on empirical
stellar libraries with the synthetic stellar libraries described in
$\S$\ref{s:sspec}.  We explore the dependence of a suite of spectral
indices on stellar age, abundance pattern, total metallicity, and IMF,
focusing on indices governed by iron, sodium, calcium, carbon, and
magnesium.  In particular, we explore variations in age between
$3-13.5$ Gyr, IMFs between a bottom-light, Chabrier, Salpeter, and
$x=3$, $\alpha-$enhancement between $0.0<$[$\alpha$/Fe]$<0.2$, and
individual elemental abundance variation.  Notice that the abundance
variations of calcium, carbon, sodium, and $\alpha-$enhancement were
implemented at fixed [Fe/H] and so [e/H]$\equiv$[e/Fe] where ``e''
refers to generic elemental abundance.  This also implies that the
total metallicity, $Z$, varies from model to model.  The red spectra
are sensitive to age in the range we consider ($3-13.5$ Gyr) primarily
because of the increased contribution of M giants to the integrated
light at younger ages.  The hotter main sequence turn-off point at
younger ages has a larger impact on the blue spectrum.

\subsubsection{Iron}

The left panel of Figure \ref{fig:FeNa} shows iron-sensitive spectral
indices.  The composite index $\langle$Fe$\rangle\equiv
0.5($FeI$0.52+$FeI$0.53)$ was defined by \citet{Trager00} and is used
extensively in the literature.  The figure shows the response of these
indices to changes in the IMF, stellar age, and abundance pattern.
The iron abundance, [Fe/H], varies by $\pm0.3$ dex.  We also incude
measurement uncertainties on each index assuming a signal-to-noise
ratio per angstrom of $S/N=200$ \AA$^{-1}$.

An important feature of these index-index plots is that the effects of
the IMF are nearly orthogonal to both age and abundance effects.  The
insensitivity of FeH to ages >5 Gyr for a Chabrier IMF found here is
in agreement with previous work \citep{Couture93, Schiavon00}.  The
response of the blue indices (along the y-axes) to age, [$\alpha$/Fe],
and [Fe/H] are well-known \citep[e.g.,][]{Schiavon07}.  In particular,
the FeI0.44 index weakens with increasing [$\alpha$/Fe] due to the
increased presence of magnesium and calcium lines in the index
pseudocontinuum \citep {Schiavon07}.

The insensitivity of the FeH index to [$\alpha$/Fe] is a key result of
this paper. This result is at first glance surprising, given the fact
that TiO absorption in M giants coincides with the index bandpass for
FeH (see Figure \ref{fig:starspec}), and Ti and O are both
$\alpha-$elements.  In fact, the strength of the TiO feature at
0.99$\mu m$ is almost entirely insensitive to 0<[$\alpha$/Fe]$<0.2$
because of two unrelated effects.  First, the atmospheric structure
changes with increased [$\alpha$/Fe] in such a way so as to
significantly diminish the effect of increased abundance of TiO
on the TiO linestrength.  This is confirmed by computing a set of
synthetic spectra where the atmospheres were computed at solar
abundances but the spectra were created with [$\alpha$/Fe]$=0.2$. In
this case the TiO features increase with $\alpha-$enhancement. The
second effect causing the FeH0.99 index to be insensitive to the TiO
abundance is the fact that the TiO feature at 0.99$\mu m$ sits atop a
much larger TiO feature whose band-head is at $0.97\mu m$ (both are
part of the $\delta$ system).  The strength of the TiO feature at
0.99$\mu m$ is determined by the ratio of the line opacity to the
pseudocontinuum, both of which scale with the abundance of TiO. For
these reasons the TiO feature at 0.99$\mu m$ feature is quite
insensitive to TiO abundance.

Notice that while the FeH0.99 index is insensitive to ages of >5 Gyr
for a Chabrier IMF, it becomes increasingly sensitive to age for
steeper IMFs. This is a consequence of the fact that at younger ages M
giants contribute more to the integrated flux, so they outshine the M
dwarfs.  This dependence of FeH0.99 on age is an important prediction
of any model that ascribes a bottom-heavy IMF to a particular stellar
population.  In other words, if a certain class of objects, say
massive ellipticals, were thought to possess a bottom-heavy IMF, then
one should find a strong correlation between the mean stellar age and
FeH0.99 strength within that population.  For a standard Chabrier IMF,
no such correlation should exist for ages $>5$ Gyr.  This prediction
is unique to the FeH0.99 index.  As will be shown in later sections,
all other IMF-sensitive lines that we consider show correlations with
age for both Chabrier and bottom-heavy IMFs.
 
The results in this section show that the combination of any blue
iron-sensitive index with the near-IR FeH index should provide an
unambiguous separation of the effects of the IMF from abundance and
age effects.

\subsubsection{Sodium}

The right panel of Figure \ref{fig:FeNa} shows sodium-sensitive
spectral indices as a function of the IMF, stellar age, and abundance
pattern.  The sodium abundance, [Na/Fe], varies by $\pm0.3$ dex.

All sodium-sensitive indices weaken with increasing [$\alpha$/Fe] due
to the fact that the pseudocontinua are strongly influenced by
$\alpha-$elements.  The NaI0.82 index decreases with decreasing age,
in agreement with previous work \citep{Schiavon00}.  The NaI0.59 index
(also known as NaD) is in theory a useful line because it responds
much more strongly to [Na/Fe] than the IMF.  Unfortunately, this index
is influenced by the interstellar medium (ISM; of the system in
question and, for systems at zero redshift, the Galaxy), and so
interpretation of this index is quite complicated.  The NaI1.14 index
is very sensitive to the IMF, as is the NaI0.82 index, although it is
also sensitive to [Na/Fe].  Both of these indices scale in the same
way with [Na/Fe] and IMF, implying that they cannot be used in
isolation to jointly constrain the IMF and sodium abundance.  The
NaI2.21 and NaI0.82 indices separate the effects of abundance and IMF
somewhat more cleanly.

For late-type giants sodium is the third most important electron
contributor at a Rosseland mean opacity of order unity, after Mg and
Al, while for late-type dwarfs it is the most important, followed by
Ca.  The abundance of sodium therefore affects the strength of many
other features through its influence on the electron pressure.  For
example, increasing the sodium abundance causes a decrease in the
abundance of CaII, which causes a decrease in the EW of CaII0.86.  The
importance of this effect should not be underestimated.  Specifically,
an increase in [Na/Fe] by 0.3 dex causes a decrease in the EW of
CaII0.86 of 1.6\%. Compare this to the decrease in CaII0.86 of 3.1\%
between a Chabrier and Salpeter IMF.  It is therefore essential to
consider not only IMF-sensitive indices such as NaI0.82 and CaII0.86
but also indices that are sensitive to [Na/Fe] in order to separate
the effects of [Na/Fe] from the IMF.  It is however worth noticing
that while an increase in [Na/Fe] can cause an increase in NaI0.82 and
a decrease in CaII0.86, mimicking the effects of a more bottom-heavy
IMF, quantitatively a very large increase in [Na/Fe] would be required
(at least 0.6 dex) to mimic an $x=3$ IMF in these indices.

In summary, unlike the iron-sensitive indices, the situation with
sodium is more complex.  Were it not for the influence of the ISM on
the NaI0.59 index, the combination of this index with NaI0.82 would
provide a powerful means for separating the IMF from other effects.
The red and near-IR sodium-sensitive indices respond strongly and
positively to both [Na/Fe] and IMF variations, making it difficult to
use these lines alone to separate IMF effects from others.  An
independent constraint of the sodium abundance would make the near-IR
sodium lines powerful IMF diagnostics.

\begin{figure*}[!t]
\center
\resizebox{3.5in}{!}{\includegraphics{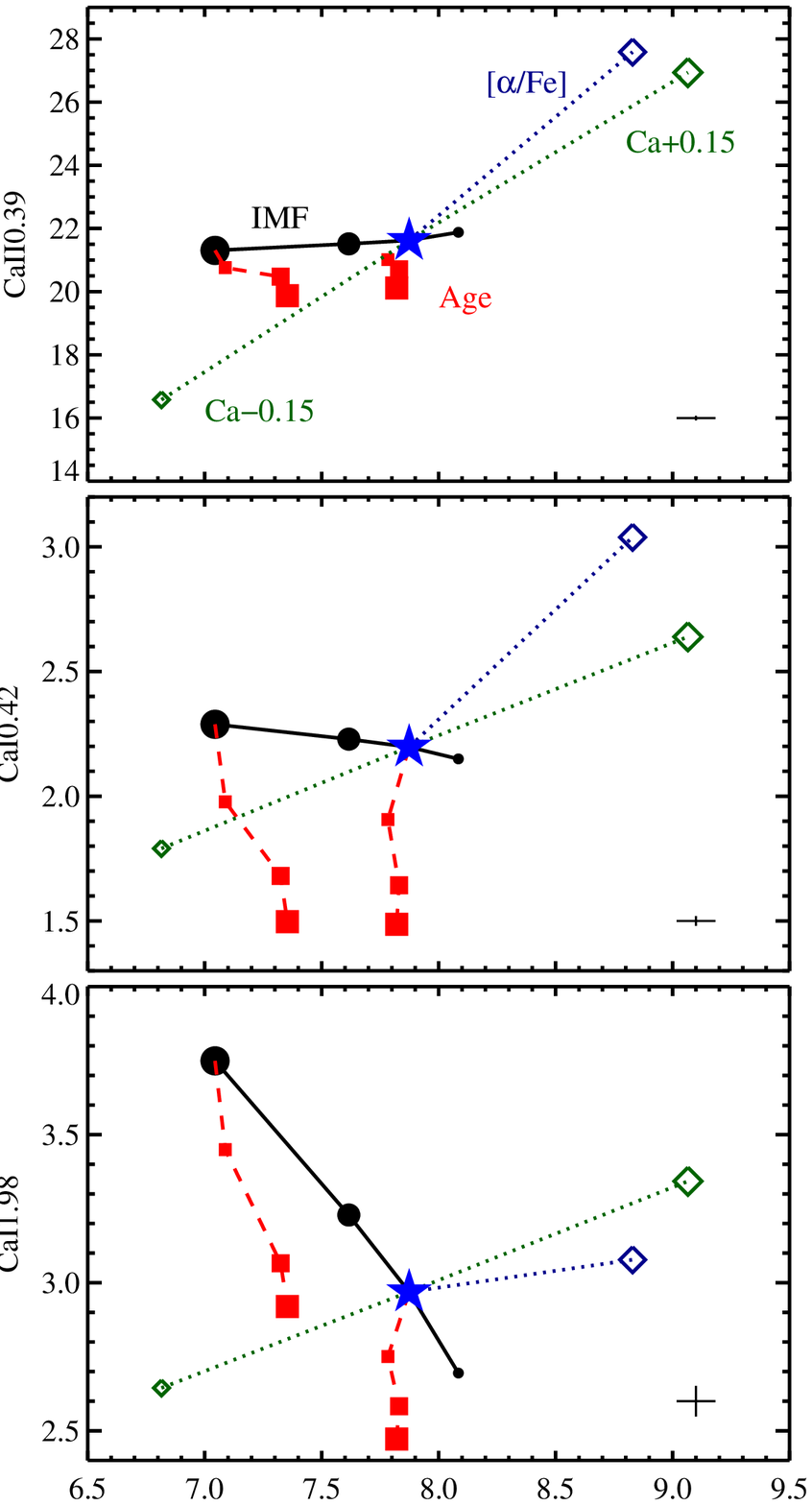}}
\resizebox{3.5in}{!}{\includegraphics{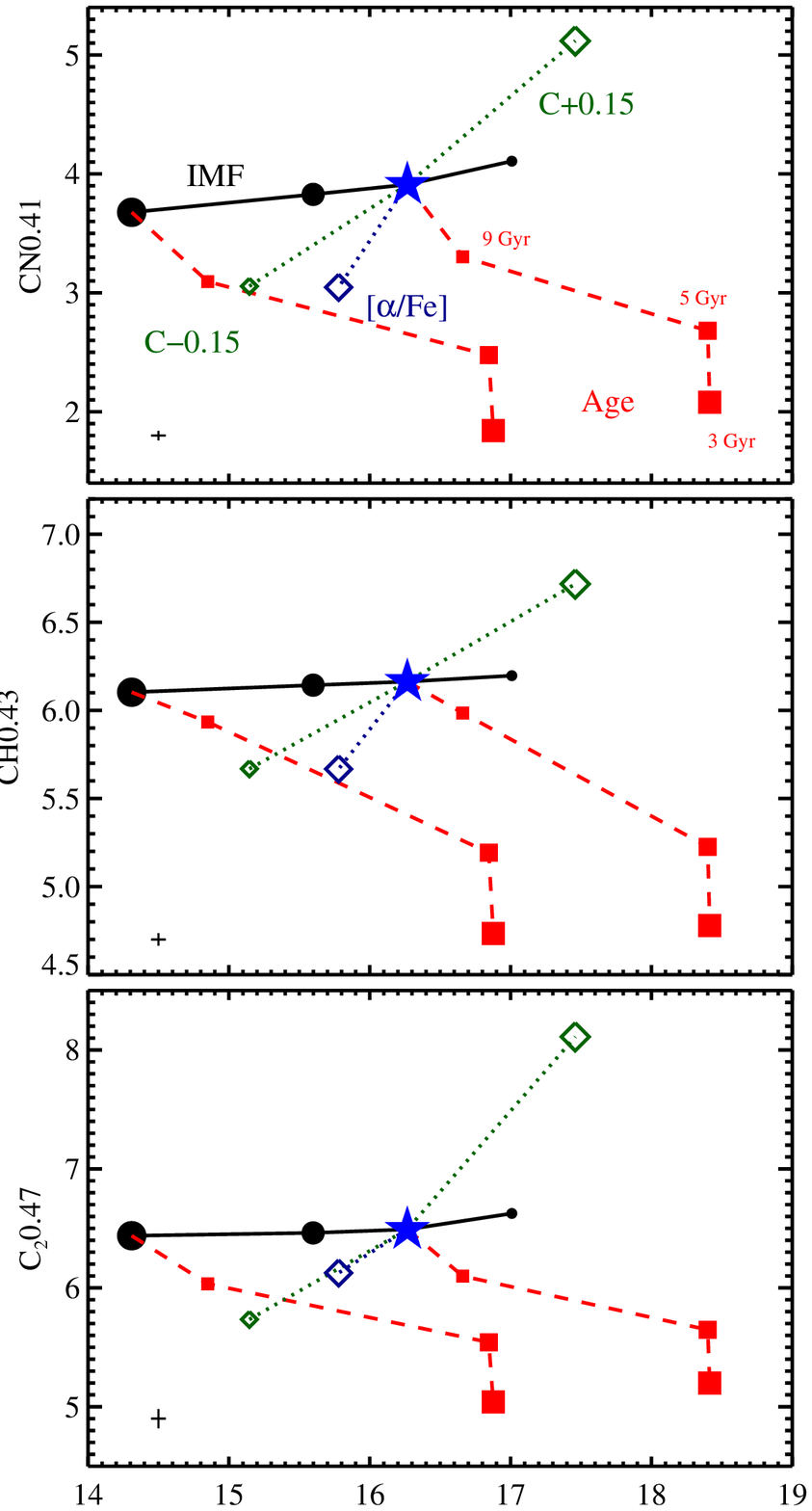}}
\vspace{0.5cm}
\caption{Same as Figure \ref{fig:FeNa}, now for calcium ({\it left
    panel}) and carbon ({\it right panel}).  The model carbon and
  calcium abundances vary by $\pm0.15$ dex.}
\label{fig:CaC}
\end{figure*}

\subsubsection{Calcium}
\label{s:ca}

The left panel of Figure \ref{fig:CaC} shows calcium-sensitive
spectral indices as a function of the IMF, stellar age, and abundance
pattern.  The calcium abundance, [Ca/Fe], varies by $\pm0.15$ dex.

The well-known IMF-sensitive CaII0.86 index (also known as the calcium
triplet or CaT) is shown here to be weakly sensitive to age, in
agreement with previous work \citep{Schiavon00, Vazdekis03}, and
strongly sensitive to both [Ca/Fe] and [$\alpha$/Fe].  Since calcium is
an $\alpha-$element, this latter fact is no surprise.

The CaII0.39 index (also known as the CaII H \& K lines) is mildly
sensitive to the IMF although it is difficult to gather this from
Figure \ref{fig:CaC} because of the large $y-$axis range.  The
difference between a Chabrier and $x=3$ IMF is only 0.2\AA.  This
index is very strongly sensitive to [Ca/Fe] and [$\alpha$/Fe]
variations \citep[see also][]{Serven05, Worthey11}.  The much larger
sensitivity of this index to abundances than the IMF suggests that it
is useful primarily as an abundance indicator.  The CaI0.42 index
behaves in a similar way, although the change in the EW induced by
abundance changes is much less dramatic.  The CaI1.98 index increases
significantly for more bottom-heavy IMFs.  This index is
proportionally much more sensitive to the IMF than abundance compared
to other calcium-sensitive indices.

The combination of the CaII0.39 or CaI0.42 indices, which are
primarily sensitive to calcium abundance, with both the CaI1.98 and
CaII0.86 indices, which respond strongly and in opposite ways to
changes in the IMF, should provide a strong constraint on the IMF.

\subsubsection{Carbon}

The right panel of Figure \ref{fig:CaC} shows carbon-sensitive
spectral indices as a function of the IMF, stellar age, and abundance
pattern.  The carbon abundance, [C/Fe], varies by $\pm0.15$ dex.  The
carbon abundance cannot increase much beyond 0.15 dex because a C/O
ratio greater than unity would produce carbon stars, which are not
expected to occur in old stellar populations \citep{Renzini81}.

The classic surface gravity-sensitive CO2.30 index is shown here to be
sensitive to the IMF, [C/Fe] abundance, age, and $\alpha-$enhancement.
The index increases with decreasing age, meaning that an increasingly
bottom-heavy IMF can be easily separated from age effects.  A decrease
in carbon abundance by 0.15 dex mimics a shift from a Chabrier IMF to
an IMF in between Salpeter and $x=3$.  CO decreases with increasing
[$\alpha$/Fe] because the atmospheric structure changes in such a way
as to decrease the CO2.30 EW.  The abundance of CO is determined by
carbon, which is less abundant than oxygen in the atmospheres of stars
comprising old stellar systems, and so CO is less sensitive to O (an
$\alpha$-element) than one may have otherwise supposed.

The blue carbon-sensitive indices CN0.41, CH0.43, and C$_2$0.47 are
all very weakly sensitive to the IMF but are strongly sensitive to
age, [C/Fe], and [$\alpha$/Fe].  These indices become weaker with
increasing $\alpha-$enhancement because increasing [$\alpha$/Fe]
increases the oxygen abundance, and, since CO has the highest
dissociation energy, more oxygen means more carbon locked up in CO,
which means less carbon available for CN and CH \citep{Serven05,
  Schiavon07}.  C$_2$0.47 is doubly sensitive to carbon abundance for
the obvious reason that it is composed of two carbon atoms.

The combination of any of the blue carbon-based lines with the CO2.30
index should provide a strong constraint on both the IMF, stellar age,
and elemental abundances.

\subsubsection{Magnesium}

\begin{figure}[!t]
\center
\resizebox{3.5in}{!}{\includegraphics{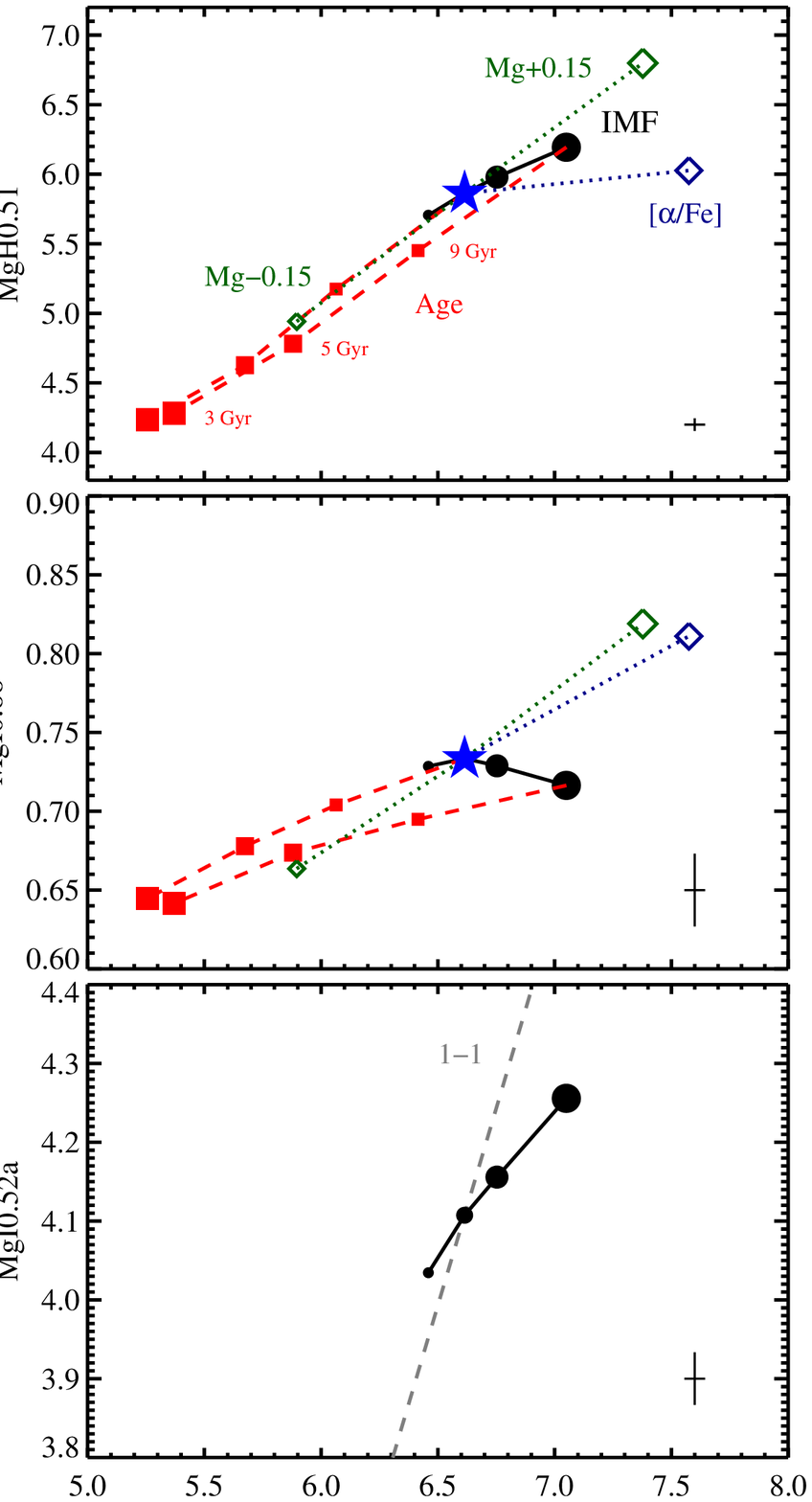}}
\vspace{0.5cm}
\caption{Same as Figure \ref{fig:FeNa}, now for magnesium.  The model
  magnesium abundances vary by $\pm0.15$ dex.  The bottom panel
  compares the classic Mg{\it b} Lick index (MgI0.52a) with our new
  MgI0.52b index.  A line with unity slope ({\it dashed line}) is
  shown for comparison.  Notice that our new index is considerably
  more IMF-sensitive than the standard Lick index.  The MgH0.51 index,
  which is analogous to the Mg$_1$ Lick index, also displays some
  sensitivity to the IMF.}
\label{fig:Mg}
\end{figure}

Magnesium spectral features have played a central role in our
understanding of the abundance patterns of old stellar systems.  For
example, the Mg{\it b} feature (consisting of a triplet of MgI lines
at $0.52\mu m$) has been used extensively to measure [$\alpha$/Fe] in
elliptical galaxies \citep[with the key assumption that Mg traces
$\alpha$; e.g.,][]{Worthey92, Trager98, Thomas05, Schiavon07}.  The
importance of magnesium in stellar population studies has motivated us
to consider this element in the context of our models.

In Figure \ref{fig:Mg} we consider four magnesium-sensitive features.
The MgH0.51 index (equivalent to the Mg$_1$ Lick index) is very
sensitive to [Mg/Fe] and moderately sensitive to age and the IMF.
Interestingly, it is insensitive to [$\alpha$/Fe]\footnote{This occurs
  because the pressure in the atmospheres of M dwarfs is lower for
  higher [$\alpha$/Fe], which results in the formation of less MgH per
  unit Mg (see $\S$\ref{s:fingerprints} for details).  This effect,
  combined with the increased abundance of Mg, results in a negligible
  change of the MgH abundance in the atmospheres of M dwarfs. The
  MgH0.51 index, which measures the strength of MgH, therefore depends
  very weakly on [$\alpha$/Fe].  When only [Mg/Fe] is increased, the
  atmospheric structure of M dwarfs is scantly affected, resulting in
  greater formation of MgH due to the increased abundance of Mg.}.
The near-IR MgI line measured by the MgI0.88 index is sensitive to
[Mg/Fe] and [$\alpha$/Fe], in agreement with previous work
\citep[e.g.,][]{Diaz89, Cenarro09}.  It is a weak feature, requiring
very high $S/N$ spectra to obtain a useful measurement, and it is
insensitive to the IMF for reasons described below.

The strength of the Mg{\it b} lines are quantified herein with two
indices: the MgI0.52a index (equivalent to the Mg{\it b} Lick index),
and the MgI0.52b index.  The latter index is new and was constructed
to be maximally sensitive to the IMF.  This is clearly demonstrated in
the bottom panel of Figure \ref{fig:Mg}.  The MgI0.52a index changes
by 0.16\AA\, between a Chabrier and $x=3$ IMF, compared to 0.46\AA\,
for the MgI0.52b index.  The reason for this difference is due to a
triplet of CrI lines centered at $\approx5208$\AA, and blended with
FeI, TiI, and MgH lines, that becomes stronger with decreasing $\teff$
and hence stronger for steeper IMFs.  The CrI blend partially overlaps
with the red pseudocontinuum bandpass of the Lick Mg{\it b} index, and
so the IMF-sensitivity of that index is greatly reduced.  Our new
index includes the CrI blend in the index definition, thereby
enhancing the effect of the IMF on the index.  We emphasize that using
the full spectrum to estimate the IMF, abundance, age, etc., rather
than spectral indices, would circumvent these complications.

There are several additional effects responsible for the IMF trends of
these indices \citep[see e.g.,][]{Schiavon04, Schiavon07}.  For
$\teff>4000$K, the strengths of MgH0.51, MgI0.52a, and MgI0.52b
increase with decreasing $\teff$ and increasing log($g$).  These
trends can be attributed primarily to the increasing strength of the
MgH and MgI spectral features.  At cooler temperatures TiO bands have
a significant, and eventually dominant, influence on these indices.
One TiO bandhead overlaps almost perfectly with the Mg{\it b} lines.
The fact that at $\teff<4000$K the Mg{\it b} Lick index becomes
primarily a measure of TiO, combined with the fact that TiO bands are
stronger in giants than in dwarfs, led \citet{Schiavon07} to conclude
that this index should not be sensitive to the IMF.  While it is true
that the Lick Mg{\it b} index is fairly weakly sensitive to the IMF,
it is primarily because of contamination by the CrI blend in the red
pseudocontinuum.  Indeed, our refined MgI0.52b index shows
significantly stronger IMF-dependence.  Although the TiO bands are not
as strong in dwarfs compared to giants, they are still sufficiently
strong to affect the Mg{\it b} region when the number of dwarfs is
increased.

Finally, the MgI0.88 index is almost completely insensitive to the IMF
in part because there is very little dependence of this feature on
log($g$) at fixed $\teff$ \citep{Cenarro09}.  This index peaks in
strength for $4000{\rm K}<\teff<4500$K, becoming weaker at both higher
and lower $\teff$. At the very coolest temperatures ($\teff<3000$K)
the dwarfs do show stronger MgI0.88 than giants (due to an
unidentified feature at $\approx8800$\AA), but the strength is
comparable to K giants, and so the addition of late M dwarfs does not
leave a signature on this index in the integrated light.

In summary, there is some sensitivity to the IMF in the Mg{\it b}
spectral region.  In this spectral region the use of indices greatly
complicates the interpretation of model variations.  Direct
manipulation of the model spectra has provided some clarity.  This
serves as yet another argument in favor of our position that model and
data should be compared directly in spectral space, rather than in
index space.  Finally, we draw attention to the fact that, while the
MgI0.52b feature displays a moderate degree of IMF sensitivity, it is
not an ideal probe of the IMF.  This is because the IMF sensitivity is
due largely to the influence of TiO, which, in the absence of any
prior on the isochrones, can be due to either dwarfs or giants.  In
other words, we could have added more M giants rather than M dwarfs,
which would have increased the strength of TiO in the model, and
thereby the strength of MgI0.52b.  We return to this point in
$\S$\ref{s:exotic}.

\begin{figure}[!t]
\center
\resizebox{3.5in}{!}{\includegraphics{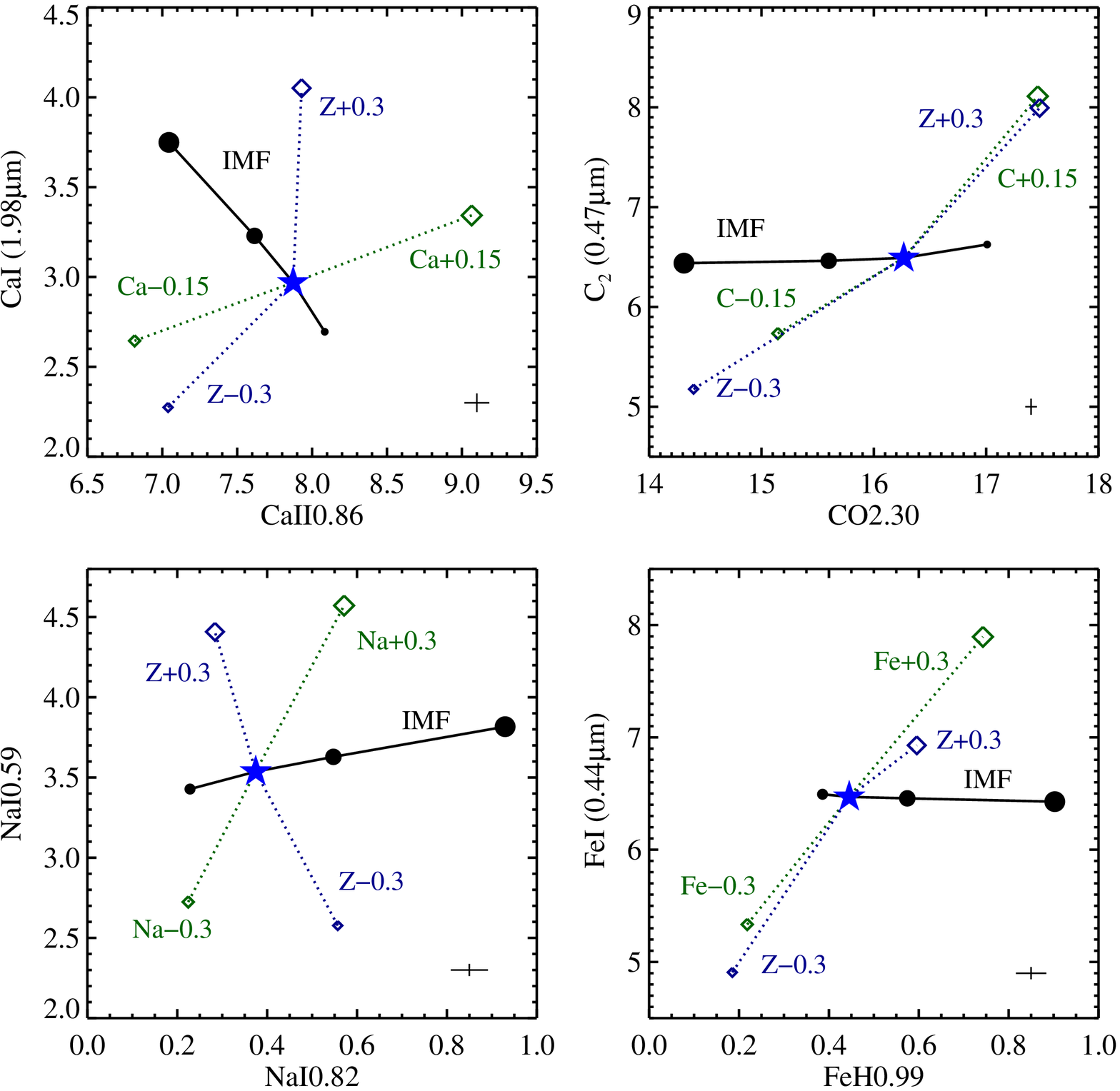}}
\vspace{0.1cm}
\caption{Effect of metallicity, $Z$, on spectral indices.  Each panel
  highlights the effect on indices dominated by a single element
  (cf. Figures 11-12).  The effect of metallicity variation of
  $\pm0.3$ dex is compared to variation in the abundance of a single
  element, and to the IMF.  The carbon and calcium abundance varies by
  $\pm0.15$ dex while the sodium and iron abundance varies by $\pm0.3$
  dex.  Symbol size increases toward higher abundances, higher
  metallicities, and steeper IMFs.  Error bars are as in Figures
  11-12.}
\vspace{0.1cm}
\label{fig:specz}
\end{figure}

\begin{figure*}[!t]
\center
\resizebox{7.4in}{!}{\includegraphics{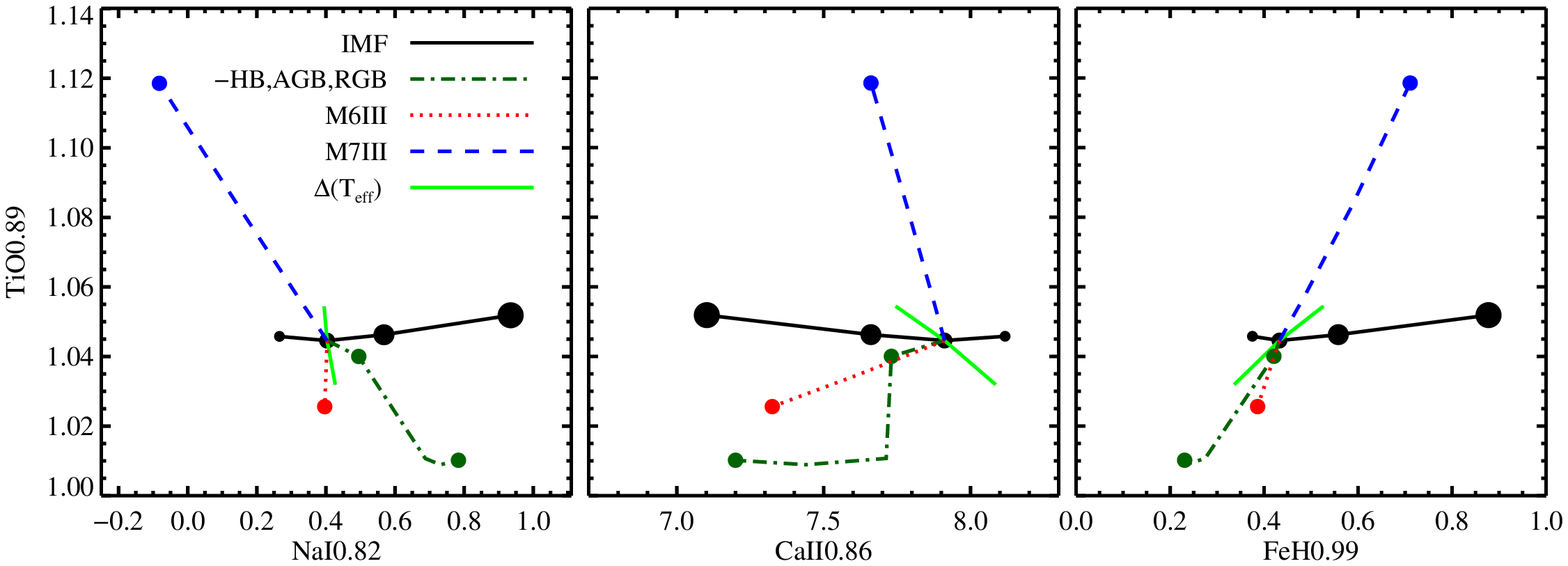}}
\vspace{0.0cm}
\caption{Response of selected spectral features to variation in the
  number and type of luminous giants and the IMF.  Models are
  constructed by removing the horizontal and asymptotic giant branches
  and also removing the most luminous RGB stars ({\it dark green
    dashed lines}), adding an increasing number of M6III ({\it red
    dashed lines}) or M7III ({\it blue dashed lines}) stars, and
  shifting the overall isochrone by $\pm50$K ({\it light green solid
    lines}).  Models with a varying IMF are also shown ({\it black
    solid lines}).  Indices along the $x-$axis are measured in EW in
  \AA\, while the Ti0.89 index is measured as a flux ratio.}
\label{fig:tio}
\end{figure*}

\subsubsection{Total metallicity}

In Figure \ref{fig:specz} we explore the effect of varying the total
metallicity, $Z$, on selected indices.  In this case the abundance
ratios are fixed to their solar values.  In this figure we have
selected a promising pair of indices that may yield IMF constraints
for each of the four elements discussed in the previous sections.  The
metallicity is varied by $\pm0.3$ dex.

The effect of metallicity variation on the indices is difficult to
interpret because the abundance of every element is changing.  In
general not only will the flux in the central bandpass change but also
the pseudocontinua used to define the index.  As an example, the
NaI0.82 index decreases with increasing metallicity because of
increased TiO absorption in the red pseudocontinuum.  In contrast, the
CO2.3 index responds to carbon abundance variation and metallicity
variation in very similar ways because the change in metallicity does
not substantially change the pseudocontinuum, and so the change in CO
abundance is the dominant effect.  The same is true for the iron
indices shown in the figure.  The CaII0.86 index decreases with
decreasing metallicity because of a change in the pseudocontinuum at
$\approx0.864\mu m$.  The independence of CaII0.86 on $Z$ for
super-solar metallicities has been noticed before by
\citet{Vazdekis03}.  These examples highlight the difficulty in using
and interpreting indices.  At this point we remind the reader that our
use of indices in this paper is for illustrative purposes only.  We do
not advocate using them when fitting models to data, in part for the
reasons discussed in this section.

The primary conclusion to be drawn from Figure \ref{fig:specz} is that
IMF variation cannot be confused with metallicity variation.  In other
words, the vectors of metallicity and IMF variation are largely
orthogonal in these index-index plots.

\subsection{Modifications to the fiducial isochrone}
\label{s:exotic}

Our goal has been to demonstrate to what extent the effects of the IMF
can be isolated from other effects on the integrated spectrum of old
stellar populations.  In the previous section we demonstrated that by
employing a variety of spectral indices one can separate the effects
of the IMF from the stellar age and abundance pattern.  Apart from
these changes to the characteristics of the stellar population, we
also need to explore changes to the assumptions that go into the
models, and in particular the precise location of the isochrones in
the HR diagram and the weights assigned to giants in the stellar
population synthesis.

We have identified three distinct isochrone-related effects that can
mimic a change in the IMF via their impact on one or more of the
IMF-sensitive indices NaI0.82, CaII0.86, and FeH0.99.  These effects
are 1) decreasing the number of luminous giants; 2) increasing the
contribution from very late M giants; 3) shifting the location of the
isochrone in the HR diagram.  We explore these effects in Figure
\ref{fig:tio} where the IMF-sensitive indices are plotted against the
TiO0.89 index, which is sensitive to both surface gravity and
temperature \citep{Carter86, Cenarro09}.  Similar index-index plots
were used by \citet{Cenarro03} to separate IMF effects from
metallicity effects.  Standard models are shown for IMFs ranging from
Chabrier to Salpeter to $x=3$ (these are the same models shown in
Figures \ref{fig:FeNa} and \ref{fig:CaC}).

We first consider the effect of decreasing the number of luminous
giants with respect to a fiducial model with an age of 13.5 Gyr, solar
metallicity, and a Chabrier IMF.  Clearly this effect should mimic the
effect of a bottom-heavy IMF to some extent since the IMF-sensitive
indices only provide a measurement of the ratio of dwarfs to giants.
The green line in Figure \ref{fig:tio} shows the effect of first
removing all HB and AGB stars, and then removing all stars at
progressively lower RGB luminosities down to $\lbol=10^2\lsun$.  This
effect clearly mimics a bottom-heavy IMF in the NaI0.82 and CaII0.86
indices, but {\it not} the FeH0.99 index.  The reason for this is
because the FeH feature is only strong in the very lowest-mass M
dwarfs, which contribute essentially no flux for a Chabrier IMF.
Removing the giants actually causes a {\it decrease} in the FeH0.99
index because this index also measures the strength of a TiO
absorption feature that occurs in late-type giants.  The TiO
absorption is a strong function of temperature, so removing the
coolest giants causes a decrease in the TiO strength, and hence the
FeH0.99 index.  From Figure \ref{fig:tio} it is clear that even for
the NaI0.82 and CaII0.86 features, one can discriminate between a
paucity of giants and a bottom-heavy IMF by considering the strength
of the TiO0.89 index.

The next effect we consider is adding more weight to very cool giants
than stellar models predict for an age of 13.5 Gyr.  Specifically, we
add increasing amounts of an M6III spectrum and an M7III spectrum to
our fiducial model.  At a maximum we include six times as many of
these giants as a 13.5 Gyr isochrone would predict.  Here again the
addition of these cool giants can in principle mimic a bottom-heavy
IMF for one or more (but not all three simultaneously) of the
IMF-sensitive features shown in Figure \ref{fig:tio}.  The addition of
cool giants mimics a bottom-heavy IMF because these stars have very
strong TiO absorption features that overlap with the FeH0.99 index and
partially overlap with the NaI0.82 feature.  The M7III actually
weakens the NaI0.82 index because it provides substantial TiO
absorption in the red pseudocontinuum of this index, thereby
decreasing the EW.  The CaII absorption at 0.86$\mu m$ weakens for the
coolest giants, explaining the trends in that index as well (this is
simply a consequence of ionization equilibrium; see Figure
\ref{fig:indteff}).  However, as with the removal of RGB stars,
consideration of these IMF-sensitive indices with the giant-sensitive
TiO0.89 feature allows one to unambiguously distinguish between the
addition of M giants and a bottom-heavy IMF.

Finally, we consider the effect of shifting the isochrone in
temperature by $\pm50$K.  Such a shift can be caused for example by a
change in [$\alpha$/Fe] by $\pm0.2$ or [Fe/H] by $\pm0.1$ dex
\citep[e.g.,][]{Dotter07, Dotter08b}.  We have modeled this effect by
running a new set of synthetic stellar spectra with $\teff$'s offset
by $\pm50$K compared to the fiducial set.  These new models were then
used differentially to quantify the effect of changing the $\teff$ of
all stars in the synthesis.  The result is shown in Figure
\ref{fig:tio}.  Qualitatively, the effect is similar to
adding/removing cool giants. The behavior of the TiO0.89 and FeH0.99
features with varying $\teff$ is due to the influence of TiO (for the
reasons noted above).  The CaII0.86 feature is sensitive to $\teff$
because of ionization equilibrium.  It is noteworthy that the NaI0.82
feature is almost completely insensitive to $\pm50$K changes to
$\teff$.  Overall, the impression from Figure \ref{fig:tio} is that
the effects of changing $\teff$ should be separable from an IMF
effect.

The preceding discussion serves to highlight another limitation of the
use of spectral indices --- as opposed to the full spectrum --- to
chart IMF variations.  Because cool stars contribute TiO to the
indices NaI0.82 and FeH0.99 that have central wavelengths and line
profiles that differ from NaI and FeH, the detailed spectral shape
around these features can also provide a strong constraint on the
contribution of cool giants to the integrated flux.

We mention a final point regarding the arbitrary addition or
subtraction of luminous giants.  In addition to the fact that stellar
evolution dictates how many such giants there should be, direct
constraints on the relative number of luminous giants come from
surface brightness fluctuations (SBFs) measured in nearby galaxies
\citep[e.g.,][]{Tonry01, Liu02, Jensen03}.  Results from SBFs do not
allow for substantial deviation from the basic predictions of stellar
evolution along the RGB and AGB.  In any event, constraints from SBFs
will provide a powerful independent test of any purported excess or
deficit of luminous giants.

We conclude that the effects considered in this section imprint
signatures not only on the IMF-sensitive features but also on the cool
giant-sensitive TiO0.89 index.  Moreover, no single effect considered
in Figure \ref{fig:tio} can simultaneously mimic an IMF effect in all
three IMF-sensitive indices.  These effects should therefore be
separable from an IMF effect given a sufficient amount of spectral
coverage.  The detailed spectral shapes of the IMF-sensitive features
will also provide a strong constraint on the contribution of cool
giants to the integrated light.

\subsection{Constraining the shape of the low-mass IMF}
\label{s:imfi}

As was alluded to in $\S$\ref{s:bsl}, the combination of multiple
surface gravity-sensitive spectral indices should provide a constraint
on the {\it shape} of the low-mass IMF in old stellar populations.
This is possible because each spectral feature responds to changes in
surface gravity and/or temperature in different ways (see Figure
\ref{fig:indteff}).  The possibility of constraining the shape of the
IMF in this way was already mentioned in \citet{Faber80}.

\begin{figure}[!t]
\center
\resizebox{3.5in}{!}{\includegraphics{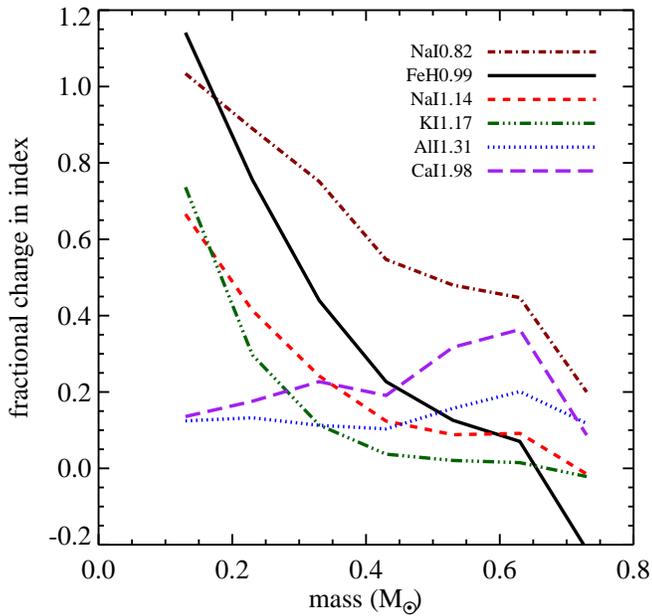}}
\caption{Sensitivity of selected spectral indices to stellar mass.
  The fractional change in each spectral index is computed by doubling
  the number of stars in each mass interval.  The index changes are
  computed with respect to a model with an age of 13.5 Gyr, solar
  metallicity, and Salpeter IMF.}
\label{fig:indmass}
\end{figure}

The sensitivity of indices to the low-mass IMF is explored in Figure
\ref{fig:indmass}.  In this figure we show the fractional change in
each index as the weight applied to each mass interval is doubled.
The change is computed with respect to a model with an age of 13.5
Gyr, Salpeter IMF, and solar metallicity.  Models constructed from the
empirical spectral libraries were used.  The FeH0.99 and KI1.17
indices are the most sensitive to very low mass stars
($\lesssim0.3\Msun$), while the CaI1.98 and AlI1.31 indices are
sensitive primarily to $0.4\Msun\lesssim M\lesssim0.7\Msun$.  The
sensitivity of the NaI0.82 index is bracketed by these two
extremes. While not shown, the giant-strong indices CaII0.86 and
CO2.30 have a much weaker dependence on mass and never exceed a
fractional change of 0.15 for any mass interval.

\begin{figure}[!t]
\center
\resizebox{3.5in}{!}{\includegraphics{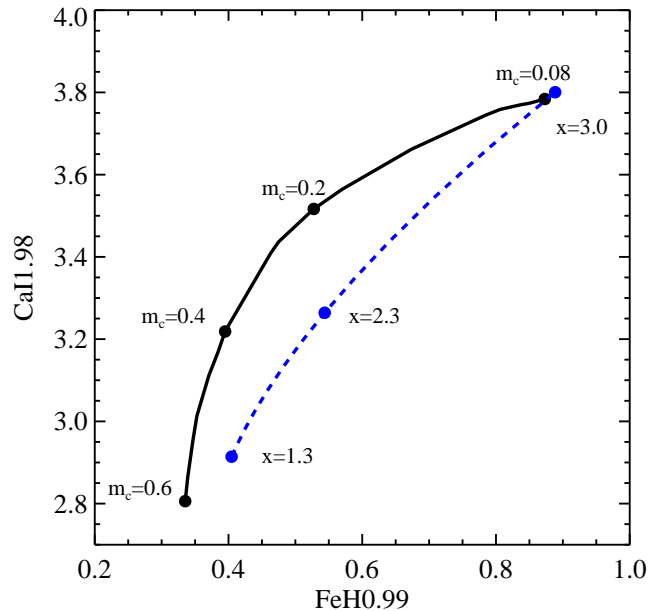}}
\caption{Index-index diagram demonstrating that the combination of
  multiple IMF-sensitive indices can constrain the shape of the
  low-mass IMF.  The CaI0.98 index is most sensitive to
  $M\approx0.5\Msun$ while the FeH0.99 index is most sensitive to
  $M<0.3\Msun$ (see Figure \ref{fig:indmass}).  The figure shows
  models at solar metallicity with an age of 13.5 Gyr.  One model
  track is shown for a power-law IMF with index $x$ that varies
  between $1.3<x<3.0$ for a fixed low-mass cutoff, $m_c=0.08\Msun$.
  The second model track is shown for a fixed power-law index of
  $x=3.0$ and varying low-mass cutoff: $0.08\Msun<m_c<0.6\Msun$.}
\vspace{0.1cm}
\label{fig:imfshape}
\end{figure}

An example showing that it is possible to measure the shape of the
low-mass IMF with these indices is shown in Figure \ref{fig:imfshape}.
In this figure we show two indices as a function of the shape of the
IMF.  The dashed lines show variation in these indices for a varying
logarithmic power-law index $x$ and fixed low-mass cutoff,
$m_c=0.08\Msun$, while the solid lines show variation in $m_c$ for a
fixed logarithmic slope $x=3.0$.  As expected, FeH0.99 responds most
quickly to changes in the low-mass cutoff while CaI1.98 responds very
weakly until $m_c>0.2\Msun$.  For $m_c>0.4\Msun$ CaI1.98 responds
strongly while FeH0.99 responds weakly, again as expected.  Changes in
the logarithmic slope result in a different relation between these
indices.

Of course, the differences induced by different IMF shapes are more
subtle than overall changes in the low-mass star content.  Constraints
on the shape of the IMF will therefore only be possible after the blue
and red indices are used to jointly constrain the detailed abundance
pattern and mean stellar age.  Once these variables are known one may
profitably use the gravity-sensitive indices to probe the detailed
shape of the low-mass IMF in the integrated light of old stellar
populations.

\subsection{Practicalities: Measuring the IMF with real data}

In this section we comment on several practical aspects regarding the
measurement of IMF-sensitive spectral features.  We focus on the
effect of velocity broadening, the required $S/N$, and the sky
subtraction and flux calibration.

\begin{figure}[!t]
\center
\resizebox{3.5in}{!}{\includegraphics{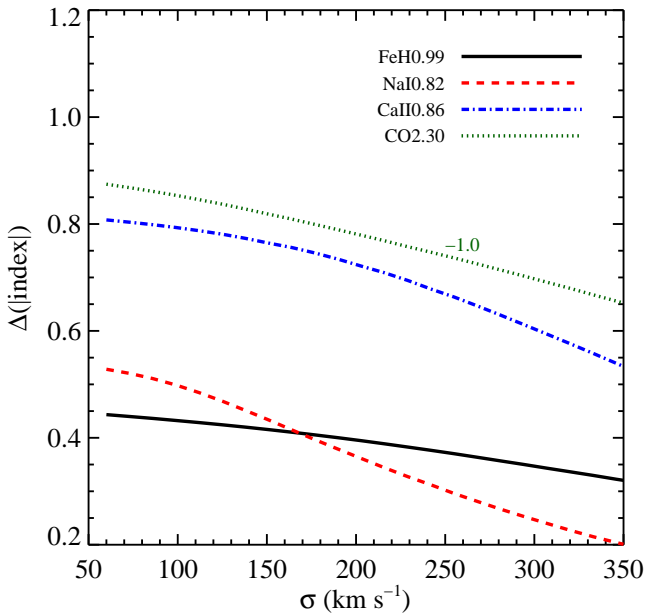}}
\caption{Absolute change in selected indices between a Chabrier and
  $x=3$ IMF as a function of velocity dispersion.  Indices are
  measured in EW with units of \AA.  The CO2.30 index has been shifted
  downward by 1\AA\, for clarity.}
\vspace{0.1cm}
\label{fig:vel}
\end{figure}

In the foregoing sections we presented results derived from spectra at
their native resolution ($R\approx2000$, or $\sigma\approx64\kms$).
Figure \ref{fig:vel} shows the extent to which the IMF sensitivity of
selected spectral features is affected by increasing the velocity
dispersion (due either to lower resolution data or an intrinsically
higher velocity dispersion population).  The figure shows the absolute
change in an index between a Chabrier and $x=3$ IMF as a function of
velocity dispersion.  Not surprisingly, the difference decreases
toward higher $\sigma$, and the effect is stronger for the CaII0.86
and NaI0.82 indices, which are relatively narrow features \citep[see
also][]{Vazdekis03}.  The FeH0.99 index is relatively insensitive to
changes in $\sigma$ because this index is dominated by broad molecular
features (FeH in dwarfs and TiO in giants).  It is reassuring to note
that while the signal is somewhat lower at high dispersion, the
differences between a bottom-heavy and Chabrier IMF are still
substantial even at $\sigma=300\kms$ \citep[see also][]{vanDokkum10}.

What is less obvious from this figure is how the detailed spectral
shape around these features changes as the velocity dispersion is
increased.  Most notable is the fact that the NaI0.82 feature is
resolved as a doublet at low dispersion, but is blended into a single
feature at $\sigma>200\kms$.  This is an example where spectral
indices neglect important details.  In low dispersion systems,
consideration of the full spectrum around the NaI doublet at 0.82$\mu
m$ would provide much stronger constraints than mere consideration of
the NaI0.82 index \citep[see Figure \ref{fig:gcspec};
also][]{Boroson91, Schiavon00}.

In Figures \ref{fig:FeNa}, \ref{fig:CaC}, and \ref{fig:Mg} we have
included error bars for $S/N=200$ \AA$^{-1}$ spectra.  These error bars
demonstrate that a spectrum with this $S/N$ is sufficient to strongly
differentiate between a Chabrier and Salpeter IMF in all of the
IMF-sensitive indices.  More generally, the $S/N$ per angstrom
required to differentiate between a Chabrier and $x=3$ IMF at the
$2\sigma$ level is 11, 29, 32, and 41 for the CO2.30, NaI0.82,
FeH0.99, and CaII0.86 indices.  In order to differentiate between a
Chabrier and Salpeter IMF at the same level of significance, the
required $S/N$ for the four indices is more demanding: 32, 105, 105
and 130.

It is difficult to make general comments regarding the level of flux
calibration and sky subtraction necessary to measure the IMF, so we
limit the following discussion to some simple statements.  Absolute
flux calibration is unnecessary for the measurement of narrow spectral
features.  Moreover, relative flux calibration is only necessary over
$\sim100$\AA\, wavelength intervals, owing to the necessity of
sampling not only the feature of interest but some continuum blueward
and redward of the feature (see Table 1).  We have experimented with
multiplying the model spectra by polynomials in order to assess the
required level of relative flux calibration.  Broadly speaking,
polynomials with 10\% amplitude variation over $\sim100$\AA\,
intervals result in $1-10$\% changes to the indices.  Of course,
pathological polynomials, where a feature in the polynomial coincides
with the feature of interest, result in much stronger changes to the
indices.  When strong sky lines coincide with a feature of interest it
will be very difficult, if not impossible, to reach the required $S/N$
in the reduced spectrum.  Stacking spectra with different redshifts in
the restframe limits (or even eliminates) the influence of sky
emission and absorption \citep[see $\S$\ref{s:data}, and
e.g.,][]{vanDokkum10,vanDokkum11}.

\section{Discussion}
\label{s:disc}

\subsection{What can be measured from integrated light}

Our primary goal has been to investigate the extent to which the
low-mass IMF can be directly constrained from integrated light
spectra.  We have demonstrated that the effect of the IMF on the red
and near-IR spectra of old stellar populations is subtle but
measurable and the effect becomes quite strong for IMFs steeper than
the disk of the Galaxy.  The only limitation of our approach is that
the stellar population must be dominated by old ($\gtrsim5$ Gyr)
stars.  Younger stars outshine the faint low mass stars by many orders
of magnitude, rendering low-mass IMF measurements impossible.  In
addition, different spectral features respond to changes in the IMF in
different ways, suggesting that the shape of the low-mass IMF can also
be directly constrained from moderate resolution ($R\approx2000$)
spectra of distant galaxies.  Based on the results shown in
$\S$\ref{s:imfi} we expect that the shape of the IMF can be
constrained in distant galaxies to at least $0.2\Msun$.  We have
presented many of the key results of this paper using spectral
indices.  We emphasize that this was merely for clarity --- when
actually fitting models to data we advocate using the spectra
directly for the reasons highlighted throughout this paper.

These results signal an important shift in stellar population studies.
With the models presented in this paper we should be able to, for the
first time, directly measure the IMF in distant galaxies, which means
that we will be able to directly measure their total {\it stellar
  mass} (not including remnants).  This measurement should be
possible, in principle, out to redshifts of order unity for old
stellar populations.  At higher redshifts the age of the Universe sets
an upper limit to the age of stellar systems that is younger than the
applicability of this technique (e.g., $\gtrsim5$ Gyr).

We have also demonstrated that the effect of the IMF in integrated
light can be separated from elemental abundance effects.  A powerful
approach is to combine blue and red spectra because the faint M dwarfs
of interest contribute approximately ten times less flux in the blue
compared to the red.  The blue spectrum is therefore much less
IMF-sensitive than the red.  Combining IMF-sensitive spectral features
in the red with IMF-insensitive blue features {\it dominated by the
  same element} (e.g., FeI0.52 and FeH0.99) is the most promising
means to separate IMF and abundance effects \citep[see also][who make
a similar point]{Vazdekis03, Cenarro04}.  The weak IMF-dependence in
the blue spectral features, especially in the Lick-based indices
explored in this work, suggests that previous work on constraining the
mean stellar age and abundance patterns --- based almost exclusively
on blue spectra and indices --- are relatively robust against changes
to the IMF.

We highlight the Wing-Ford band, denoted herein as FeH0.99, as the
most promising single feature for constraining the low-mass IMF. It is
insensitive to age for ages $>5$ Gyr and insensitive to
$\alpha-$enhancement.  It depends on the iron abundance, which can be
measured accurately and independently in the blue.  It also depends
only weakly on velocity dispersion because it is intrinsically broad.
It is therefore an ideal indicator of the dwarf-to-giant ratio in old
stellar populations.

All of the model results in this paper make the implicit assumption
that the IMF is fully populated.  This assumption is valid for systems
with minimum masses in excess of $10^{5-6}\Msun$, depending on
wavelength and age \citep{Lancon00, Cervino04, Popescu10}.  For
galaxies and massive globular clusters, our assumption is therefore
valid, while for less massive systems stochastic effects will need to
be modeled.

\subsection{Previous work}

It has been known since at least the work of \citet{Spinrad62} that
one could in principle measure the contribution of faint M dwarfs to
the integrated light of old stellar populations.  As noted in the
introduction, previous observational work suffered from a number of
technical obstacles including poor detector technology.  Poor spectral
resolution was also a problem.  In an important early paper,
\citet{Carter86} attempted to measure the strength of the NaI0.82
feature in massive ellipticals, but they could not separate the NaI
feature from the adjacent TiO feature, which is strong in giants,
greatly complicating the interpretation \citep[see also][]{Delisle92}.
This is no longer a limitation with current instruments.

It is also worth noting that the scope of the discussion has changed
markedly since the early papers of \citet{Spinrad72}, \citet{Cohen79},
and \citet{Faber80}.  In that early work the question was whether or
not M dwarfs {\it dominated}, or at least contributed significantly,
to the integrated light.  \citet{Spinrad72} claimed, for example, that
dwarfs contributed $\sim30$\% to the $V-$band light of old
populations.  This would imply a mass-to-light ratio of
$M/L\approx40$, which we know now is strongly ruled out on dynamical
grounds \citep[e.g.,][]{Cappellari06}.  It is now generally agreed
that dwarfs do not contribute substantially to the integrated light.
The debate today centers on whether dwarfs contribute $\lesssim1$\% or
several percent to the integrated light.  This difference is quite
subtle but of fundamental importance not only for questions regarding
the universality of the IMF but also the relative amount of dark and
luminous matter in the centers of galaxies.

Early models attempted to construct integrated light spectra from
empirical stellar spectra \citep[e.g.,][]{Carter86}.  Among the
limitations was again the quality of the data, and also the fact that
the stellar libraries contained few M dwarfs.  \citet{Schiavon00} were
the first to create fully synthetic population synthesis predictions
in the near-IR with the aim of investigating the effects of the IMF,
population age, and metallicity on the NaI, CaII, and FeH features.
The state of population synthesis modeling has not changed
qualitatively in the intervening decade.  The most important
improvements between this paper and \citet{Schiavon00} are 1) the more
comprehensive analysis of the blue through near-IR spectra as a
function of IMF, age, and individual elemental abundances, 2) the use
of empirical spectral libraries for the fiducial models, with the
synthetic models being used only differentially, and 3) important
updates to the model atmosphere and spectra calculations, including
the addition of the H$_2$O linelist and improvements in other existing
linelists such as TiO and FeH.

More recently, \citet{Cenarro03} concluded that the IMF varies with
velocity dispersion and [Fe/H] based on analysis of the CaII triplet,
TiO, and MgI lines.  That work lacked models with variable abundance
ratios, and so the authors could not reliably separate IMF effects
from abundance effects.  Indeed, interpretation of calcium line
strengths have led to many contradictory claims over the years,
without a clear resolution \citep{Thomas03b, Falcon-Barroso03,
  Cenarro04, Worthey11}.  The new models presented in the present work
should be able to resolve the calcium puzzle.

\subsection{van Dokkum \& Conroy (2010)}

In \citet{vanDokkum10} we used the NaI0.82 and FeH0.99 spectral
features to probe the low-mass IMF in eight massive ellipticals.
Based on high signal-to-noise, $R\sim2000$ spectra taken with the Keck
telescope, we concluded that these galaxies had on average a much more
bottom-heavy IMF compared to the Galaxy; an IMF between Salpeter and
$x=3$ was preferred, although a quantitative assessment was not
provided in that work.  With the model results presented in this paper
we are now in a position to reassess those conclusions.

The result of a bottom-heavy IMF in massive ellipticals was based on a
preliminary version of the population synthesis model presented
herein.  In particular, the base isochrones and empirical IRTF spectra
are essentially identical between that work and the present paper.
The primary uncertainty in our previous work was the uncertain impact
of elemental abundance variations and stellar population age effects
on the NaI0.82 and FeH0.99 indices.  This concern was particularly
acute since it is known that the galaxies we targeted show
$\alpha-$enhanced abundance patterns \citep[in subsequent work we
compared these massive elliptical spectra to metal-rich globular
clusters in M31 that have similar abundance patterns and again found
evidence for a bottom-heavy IMF in the ellipticals; see][for
details]{vanDokkum11}.  In fact, one of the motivations for the
present work was to assess the effect of abundance variations on our
previous results.

In light of these concerns, the most important results from the
present work are that the FeH0.99 index is essentially insensitive to
[$\alpha$/Fe] and the NaI0.82 index is anti-correlated with
[$\alpha$/Fe].  Thus, the construction of models more applicable to
massive ellipticals, where [$\alpha$/Fe]$\sim0.2$ and [Fe/H]$\sim0.0$,
will produce FeH0.99 and NaI0.82 strengths that {\it still demand a
  bottom-heavy IMF} in our sample of massive ellipticals.  In order to
explain the strengths of these features as being due to abundance
effects would require an average iron abundance in excess of solar
([Fe/H]$\sim0.15$) and an extremely high average sodium abundance
([Na/Fe]$>0.5$).  The former would result in strong disagreement with
the blue FeI lines.  The latter is not directly constrained for
massive ellipticals, but we note that in the bulge of the Galaxy,
which is $\alpha-$enhanced like the massive ellipticals, [Na/Fe] is
never greater than 0.3 dex and has an average of [Na/Fe]=0.2
\citep{Fulbright07}.  The only stars known to have very high sodium
abundances ([Na/Fe]>0.5) are found exclusively within globular
clusters \citep{Gratton04}.

Beyond confirming the somewhat qualitative analysis in our previous
work, we aim to quantify the shape of the IMF and better constrain the
number of low mass stars in massive elliptical galaxies. This requires
a quantitative analysis of the blue and red spectra of our sample of
massive ellipticals based on the models presented in this paper and is
the subject of ongoing work.

\subsection{On the fingerprints of M dwarfs in integrated light}
\label{s:fingerprints}

The fundamental idea underpinning this paper is the fact that certain
spectral features betray the presence of faint M dwarfs in integrated
light spectra.  This idea is based partly on the fact that certain
spectral features depend strongly on surface gravity at fixed
effective temperature.  Although this fact is well known both
empirically and via inspection of model spectra, we have been
surprised by the lack of discussion in the literature regarding its
physical origin.  We have therefore endeavored to understand these
surface gravity trends.  What follows is a brief sketch of the
basic physics involved; a more quantitative investigation is left for
future work.

Careful inspection of Figure \ref{fig:specrat} reveals that every
atomic feature that is strong in dwarfs is a neutral metal (e.g., NaI,
KI, CaI, AlI), while every atomic feature that is strong in giants is
a singly-ionized metal (specifically, CaII).  This is a simple
consequence of ionization equilibrium (i.e., the Saha equation): as
the electron pressure increases at fixed temperature (due to the
increased surface gravity), the ratio of singly-ionized to neutral
species of the same atom decreases.  We have confirmed this directly
with the model atmospheres: for stars with $\teff=3500$ K, dwarfs have
a higher relative abundance of NaI and a lower relative abundance of
CaII compared to the giants.

The situation with molecules is similar to that of the atoms.
Molecular dissociation equilibrium \citep{Russell34, Tsuji73} implies
that an increase in the partial pressures of the atoms and molecules
favors the side of a given chemical reaction that results in fewer
moles of gas.  For example, the formation of H$_2$O is favored because
its formation results in fewer moles of gas compared to its atomic
constituents.  This is also confirmed upon inspection of the model
atmospheres: at a fixed effective temperature, molecules are in
greater abundance with respect to hydrogen for the dwarfs compared to
the giants.

An important exception to this rule is CO.  It has the highest
dissociation energy of any known molecule and therefore its formation
is highly favored over other molecules that might compete for free
carbon and oxygen.  Its formation is so favored, in fact, that in the
atmospheres of cool giants essentially all the carbon is locked in CO.
This is important because it means that as the pressure is increased
moving from a giant to a dwarf, the relative abundance of CO does {\it
  not} increase in the manner outlined above; all the carbon is
already locked up in the most tightly bound molecule.  This probably
explains why CO features behave differentially from other features
dominated by molecules.

Of course, consideration of the abundances of atoms and molecules is
not sufficient to determine the qualitative behavior of spectral
features.  The strength of any spectral line is determined by the
ratio of line opacity to continuum opacity.  For example, the
abundance of TiO is higher in dwarfs than in giants, indicating a
higher line opacity, but the strength of TiO features is frequently
much stronger in giants than in dwarfs.  The difference must be due to
the variation in the continuum opacity between dwarfs and giants.  The
situation is especially complex for M stars, as the near-IR continuum
is largely determined by the H$_2$O opacity.  Consideration of these
effects is beyond the scope of the present paper.  The basic point we
wish to make here is that it should not be surprising that the neutral
metals generically behave oppositely to the singly-ionized metals, and
that CO should behave differently from all other molecules in
late-type stars.

For certain features there is yet another factor determining their
sensitivity to M dwarfs.  From the isochrones shown in Figure
\ref{fig:iso1} it is clear that the faintest M dwarfs are
substantially cooler than the faintest M giants for an age of 13.5 Gyr
(2000 K vs. 3000 K).  It is also evident from Figure \ref{fig:indteff}
that features such as NaI0.82, FeH0.99, and KI1.17 become {\it much}
stronger at $\teff<3000$ K.  This is an important point because it
means that at least some IMF-sensitive features are IMF-sensitive both
because they are sensitive to surface gravity and because they vary
strongly with effective temperature amongst late-type stars.

\section{Summary}
\label{s:sum}

We now summarize our main conclusions.

\begin{itemize}

\item Low-mass stars impart a unique signature on the integrated
  spectrum of old stellar populations.  Spectral features sensitive to
  low-mass stars reside in the red and near-IR and arise from lines of
  various elements and molecules (including Ca, Na, K, Al, CO, and
  FeH).  The low-mass IMF can therefore be inferred directly from
  red/near-IR spectra of old populations despite the fact that
  low-mass stars comprise only a few percent of the bolometric
  luminosity.

\item These IMF-sensitive features also vary with elemental abundance
  patterns and some vary with stellar age.  Fortunately, the
  combination of IMF-sensitive features in the red with
  IMF-insensitive features in the blue provides a joint constraint on
  the IMF, stellar age, metallicity, and abundance pattern.  The FeH
  feature at 0.99$\mu m$ is particularly promising because it is
  insensitive to $\alpha$-enhancement and stellar ages of $5-13.5$
  Gyr.  The triplet of CaII features near 0.86$\mu m$ are also
  promising if considered in conjunction the CaI doublet at 1.98$\mu
  m$ since these features respond strongly and with opposite sign to
  variations in the IMF.

\item It is difficult to mimic IMF effects with other exotic
  modifications to standard population synthesis, such as an arbitrary
  increase or decrease in the number of RGB, HB, and/or AGB stars.
  The strength of the TiO band-head at 0.89$\mu m$ provides a strong
  constraint on the contribution of very late M giants to the
  integrated spectrum.

\item Each of the IMF-sensitive features explored in this work is
  sensitive to a different regime of the low-mass IMF.  It should
  therefore be possible to directly constrain the {\it shape} of the
  low-mass IMF from integrated spectra of old stellar populations.

\end{itemize}

It has been appreciated for several decades that it is possible to
constrain the mean stellar age, metallicity, and abundances of at
least some elements from the integrated light spectrum of old stellar
populations.  The models presented in this work demonstrate that the
low-mass IMF should be added to this list of physical properties
to be constrained directly by the data.

The models presented in this paper will be made available upon request.


\acknowledgments 

We acknowledge fruitful conversations with Aaron Dotter, Marijn Franx,
Bob Kurucz, and Ricardo Schiavon.  We also thank Bob Kurucz for his
expert advice regarding synthetic stellar spectra and for compiling
the FeH linelist specifically for this project.  Nelson Caldwell is
thanked for providing his blue M31 GC spectra.  We thank the referee
for constructive comments. The synthetic spectral libraries were
computed on the Odyssey cluster supported by the FAS Science Division
Research Computing Group at Harvard University.


\end{document}